# Supplementary Results of a Comparative Syntactic and Semantic Study of Terms for Software Testing Glossaries

**Luis Olsina**[1], **Philip Lew**[2] and **Guido Tebes**[1]
[1]GIDIS_Web, Facultad de Ingeniería, UNLPam, General Pico, LP, Argentina
[2]XBOSoft, Woodbridge, VA, USA
{guido_tebes, olsinal}@ing.unlpam.edu.ar;
philiplew@gmail.com

**Abstract.** This preprint specifies supplementary material and the results of a comparative, syntactic, and semantic study of terms for three software testing glossaries. The three software testing glossaries are: the ISO 29119-1 "Concepts and definitions" standard, the Standard Glossary of Terms used in Software Testing (version 3.5) by the International Software Testing Qualifications Board (ISTQB), and the glossary provided by the Test Maturity Model integration (TMMi) Foundation. First, we categorized the glossaries terms by analyzing their given semantics into 8 categories, namely: test project/strategy/organization; testing process/activity; test goal/requirement/entity; test work product; testing method/technique; testing agent/role/tool; other terms somewhat related to test; and terms beyond the testing domain. Also, we designed a set of metrics that serve to extract and produce data that support the analysis of syntactic and semantic similarities and discrepancies, and ultimately the analysis of consistencies and inconsistencies between glossaries' terms. It is important to remark that in this work the whole analysis is focused on glossaries terms whose names end with the word "testing".

## Preprint Roadmap

This Section provides an overview of how this preprint is organized into content-specific appendices, which are supplementary material to a conference article [4]. There are six appendices, namely:

- Appendix I specifies the 8 defined categories for the glossary terms.
- Appendix II presents the classification of the ISO 29119-1 glossary terms [1] into the 8 defined categories in addition to all terms and their definitions and synonyms taken from the source.
- Appendix III contains the classification of the TMMi glossary terms [2] into the 8 defined categories in addition to all terms and their definitions and synonyms taken from the source.
- Appendix IV documents the classification of the ISTQB glossary terms [3] into the 8 defined categories in addition to all terms and their definitions and synonyms taken from the source.
- Appendix V shows the quantities related to the number of terms for each category in the glossaries.
- Appendix VI contains a set of tables that indicates the frequency of each glossary term ending with the word "testing" regarding its syntactic similarity, in addition to its category with regard to the given semantic by the sources.

## References

[1] *Software and systems engineering – Software Testing – Part 1: Concepts and definitions*, ISO/IEC/IEEE 29119-1, 2013.

[2] *Test Maturity Model Integration (TMMi®) - Guidelines for Test Process Improvement*, TMMi Foundation, release 1.2, 2018.

[3] *Standard Glossary of Terms used in Software Testing*, International Software Testing Qualifications Board (ISTQB®), version 3.5, 2021. [Online]. Available: https://www.istqb.org/.

[4] Olsina L., Lew P., Tebes G.: Analyzing Quality Issues from Software Testing Glossaries used in Academia and Industry. To appear in Springer Nature Switzerland AG book, CCIS: 15th Int'l Conference on the Quality of Information and Communication Technology, Vallecillo A., and Visser J. (Eds.): QUATIC 2022, September 12-14, Talavera de la Reina, Spain, pp. 1-15, 2022.

# Appendix I: Terminological Categories for Terms in Software Testing

*This Appendix contains a table with the description of each terminological category used in this work.*

| Categ.ID | Terminological Category name |
|---|---|
| C1 | Test Project-, Strategy-, Organizational Test-related Terms |
| C2 | Testing Work Process-, Activity-related Terms |
| C3 | Test Goal-, Requirements-, Entity-related Terms |
| C4 | Test Work Product-related Terms (e.g. Artifact, Report, Result, Specification) |
| C5 | Testing Method-, Technique-, Procedure-, Rule-related Terms |
| C6 | Testing Agent-, Role-, Tool-related Terms |
| C7 | Other Terms somewhat related to Test (e.g., Anomaly, Defect, etc.) |
| C8 | Terms beyond the Test Domain related to Quality or Software Engineering |

# Appendix II: ISO 29119-1 glossary terms and their classification

*This Appendix contains all terms, synonyms and definitions of the ISO 29119-1 software testing glossary. Also, we included the classification of all terms considering the 8 categories mentioned in Appendix I.*

| **ISO 29119-1 Glossary** | **ISO 29119-1 Terms Classified into 8 Categories or Conceptual Blocks for Testing.** Note that we include the synonyms of the terms in parentheses and not in bold. |
|---|---|
| **Cat#1 (C1): Test Project-, Strategy-, Organizational Test-related Terms** | |
| **test strategy; Organizational Test Process; Organizational Test Policy; organizational test specification; Organizational Test Strategy; suspension criteria; test practice** (test approach) | |
| **Cat#2 (C2): Testing Work Process-, Activity-related Terms** | |
| **accessibility testing; backup and recovery testing; capacity testing; compatibility testing; dynamic testing; endurance testing; equivalence partitioning; error guessing; exploratory testing; installability testing; load testing; maintainability testing; performance testing; portability testing; procedure testing; regression testing; reliability testing; retesting** (confirmation testing); **risk-based testing; scenario testing; scripted testing; security testing; specification-based testing** (black-box testing; closed box testing); **statement testing; static testing; stress testing; structure-based testing** (structural testing; glass-box testing; white box testing); **Test Completion Process; Test Design and Implementation Process; test design technique** (test technique); **Test Environment Set-up Process; test execution; Test Execution Process; Test Incident Reporting Process; test level** (test phase); **test management; Test Management Process; Test Monitoring and Control Process; Test Planning Process; test process; test sub-process; test type; testing; unscripted testing; volume testing** | |
| **Cat#3 (C3): Test Goal-, Requirements-, Entity-related Terms** | |
| **feature set; test condition** (test requirement); **test environment; Test Environment Requirements; test item** (test object) | |
| **Cat#4 (C4): Test Work Product-related Terms (e.g. Artifacts, Reports, Results, Specifications)** | |
| **actual results; expected results; Incident Report; test basis; test case; Test Case Specification; Test Completion Report** (Test Summary Report); **equivalence partition coverage; statement coverage; test coverage; test coverage item** (coverage item); **test data; Test Data Readiness Report; Test Design Specification; test environment readiness report; Test Execution Log; Test Plan; Test Procedure Specification** (manual test script); **test result; test script; test set; test specification; test status report; test traceability matrix** (verification cross reference matrix; requirements test matrix; requirements verification table); **testware** | |
| **Cat#5 (C5): Testing Method-, Technique-related, Procedure-related, Rule-related Terms** | |
| **pass/fail criteria; test procedure** | |
| **Cat#6 (C6): Testing Agent-, Role-,Tool-related Terms** | |
| | |
| **Cat#7 (C7): Other Terms somewhat related to Test (e.g., anomaly, defect, etc.)** | |
| **equivalence partition; project risk; product risk** | |
| **Cat#8 (C8): Terms beyond to the Test Domain related to Quality and Software Engineering** | |
| **Decision** | |

| **Term name** synonyms | **Definition** |
|---|---|
| **actual results** | set of behaviours or conditions of a test item, or set of conditions of associated data or the test environment, observed as a result of test execution |
| **backup and recovery testing** | type of reliability testing that measures the degree to which system state can be restored from backup within specified parameters of time, cost, completeness, and accuracy in the event of failure |
| **capacity testing** | type of performance efficiency testing conducted to evaluate the level at which increasing load (of users, transactions, data storage, etc.) compromises a test item's ability to sustain required performance |
| **compatibility testing** | type of testing that measures the degree to which a test item can function satisfactorily alongside other independent products in a shared environment (co-existence), and where necessary, exchanges information with other systems or components (interoperability) |
| **decision** | types of statements in which a choice between two or more possible outcomes controls which set of actions will result |
| **dynamic testing** | testing that requires the execution of the test item |
| **endurance testing** | type of performance efficiency testing conducted to evaluate whether a test item can sustain a required load continuously for a specified period of time |
| **equivalence partition** | subset of the range of values of a variable, or set of variables, within a test item or at its interfaces such that all the values in the partition can reasonably be expected to be treated similarly by the test item (i.e. they may be considered "equivalent") by the test item |
| **equivalence partition coverage** | proportion of identified equivalence partitions of a test item that are covered by a test set<br><br>Note 1 to entry: In many cases, the identification of equivalence partitions is subjective (especially in the sub-partitioning of "invalid" partitions), so a definitive count of the number of equivalence partitions in a test item could be impossible. |
| **equivalence partitioning** | test design technique in which test cases are designed to exercise equivalence partitions by using one or more representative members of each partition |
| **error guessing** | test design technique in which test cases are derived on the basis of the tester's knowledge of past failures, or general knowledge of failure modes |
| **expected results** | observable predicted behaviour of the test item under specified conditions based on its specification or another source |
| **exploratory testing** | experience-based testing in which the tester spontaneously designs and executes tests based on the tester's existing relevant knowledge, prior exploration of the test item (including the results of previous tests), and heuristic "rules of thumb" regarding common software behaviours and types of failure |
| **feature set** | collection of items which contain the test conditions of the test item to be tested which can be collected from risks, requirements, functions, models, etc. |
| **Incident Report** | documentation of the occurrence, nature, and status of an incident |
| **installability testing** | type of portability testing conducted to evaluate whether a test item or set of test items can be installed as required in all specified environments |
| **load testing** | type of performance efficiency testing conducted to evaluate the behaviour of a test item under anticipated conditions of varying load, usually between anticipated conditions of low, typical, and peak usage |
| **maintainability testing** | test type conducted to evaluate the degree of effectiveness and efficiency with which a test item may be modified |
| **Organizational Test Policy** | an executive-level document that describes the purpose, goals, and overall scope of the testing within an organization, and which expresses why testing is performed and what it is expected to achieve Note 1 to entry: It is generally preferable to keep the Organizational Test Policy as short as possible in a given context. |
| **Organizational Test Process** | test process for developing and managing organizational test specifications |
| **organizational test specification** | document that provides information about testing for an organization, i.e. information that is not project-specific |
| **Organizational Test Strategy** | document that expresses the generic requirements for the testing to be performed on all the projects run within the organization, providing detail on how the testing is to be performed |
| **pass/fail criteria** | decision rules used to determine whether a test item, or feature of a test item, has passed or failed after testing |
| **performance testing** | type of testing conducted to evaluate the degree to which a test item accomplishes its designated functions within given constraints of time and other resources |
| **portability testing** | type of testing conducted to evaluate the ease with which a test item can be transferred from one hardware or software environment to another, including the level of modification needed for it to be executed in various types of environment |
| **procedure testing** | type of functional suitability testing conducted to evaluate whether procedural instructions for interacting with a test item or using its outputs meet user requirements and support the purpose of their use |
| **product risk** | risk that a product may be defective in some specific aspect of its function, quality, or structure |
| **project risk** | risk related to the management of a project |
| **regression testing** | testing following modifications to a test item or to its operational environment, to identify whether regression failures occur |
| **reliability testing** | type of testing conducted to evaluate the ability of a test item to perform its required functions, including evaluating the frequency with which failures occur, when it is used under stated conditions for a specified period of time |
| **retesting** confirmation testing | re-execution of test cases that previously returned a "fail" result, to evaluate the effectiveness of intervening corrective actions |
| **risk-based testing** | testing in which the management, selection, prioritisation, and use of testing activities and resources are consciously based on corresponding types and levels of analyzed risk |
| **scenario testing** | class of test design technique in which tests are designed to execute individual scenarios |
| **scripted testing** | dynamic testing in which the tester's actions are prescribed by written instructions in a test case |
| **security testing** | type of testing conducted to evaluate the degree to which a test item, and associated data and information, are protected so that unauthorized persons or systems cannot use, read, or modify them, and authorized persons or systems are not denied access to them |
| **specification-based testing** black-box testing; closed box testing | testing in which the principal test basis is the external inputs and outputs of the test item, commonly based on a specification, rather than its implementation in source code or executable software |
| **statement coverage** | percentage of the set of all executable statements of a test item that are covered by a test set |
| **statement testing** | test design technique in which test cases are constructed to force execution of individual statements in a test item |
| **static testing** | testing in which a test item is examined against a set of quality or other criteria without code being executed |
| **stress testing** | type of performance efficiency testing conducted to evaluate a test item's behaviour under conditions of loading above anticipated or specified capacity requirements, or of resource availability below minimum specified requirements |
| **structure-based testing** structural testing; glass-box testing; white box testing | dynamic testing in which the tests are derived from an examination of the structure of the test item |
| **suspension criteria** | criteria used to (temporarily) stop all or a portion of the testing activities |
| **test basis** | body of knowledge used as the basis for the design of tests and test cases |
| **test case** | set of test case preconditions, inputs (including actions, where applicable), and expected results, developed to drive the execution of a test item to meet test objectives, including correct implementation, error identification, checking quality, and other valued information |
| **Test Case Specification** | documentation of a set of one or more test cases |
| **Test Completion Process** | Test Management Process for ensuring that useful test assets are made available for later use, test environments are left in a satisfactory condition, and the results of testing are recorded and communicated to relevant stakeholders |
| **Test Completion Report** Test Summary Report | report that provides a summary of the testing that was performed |
| **test condition** test requirement | testable aspect of a component or system, such as a function, transaction, feature, quality attribute, or structural element identified as a basis for testing |
| **test coverage** | degree, expressed as a percentage, to which specified test coverage items have been exercised by a test case or test cases |
| **test coverage item**<br>coverage item | attribute or combination of attributes that is derived from one or more test conditions by using a test design technique that enables the measurement of the thoroughness of the test execution |
| **test data** | data created or selected to satisfy the input requirements for executing one or more test cases, which may be defined in the Test Plan, test case or test procedure |
| **Test Data Readiness Report** | document describing the status of each test data requirement |
| **Test Design and Implementation Process** | test process for deriving and specifying test cases and test procedures |
| **Test Design Specification** | document specifying the features to be tested and their corresponding test conditions |
| **test design technique**<br>test technique | activities, concepts, processes, and patterns used to construct a test model that is used to identify test conditions for a test item, derive corresponding test coverage items, and subsequently derive or select test cases |
| **test environment** | facilities, hardware, software, firmware, procedures, and documentation intended for or used to perform testing of software |
| **test environment readiness report** | document that describes the fulfilment of each test environment requirement |
| **Test Environment Requirements** | description of the necessary properties of the test environment |
| **Test Environment Set-up Process** | dynamic test process for establishing and maintaining a required test environment |
| **test execution** | process of running a test on the test item, producing actual result(s) |
| **Test Execution Log** | document that records details of the execution of one or more test procedures |
| **Test Execution Process** | dynamic test process for executing test procedures created in the Test Design and Implementation Process in the prepared test environment, and recording the results |
| **Test Incident Reporting Process** | dynamic test process for reporting to the relevant stakeholders issues requiring further action that were identified during the test execution process |
| **test item**<br>test object | work product that is an object of testing |
| **test level**<br>test phase | specific instantiation of a test sub-process<br><br>EXAMPLE The following are common test levels that can be instantiated as test sub-processes: component test level/sub-process, integration test level/sub-process, system test level/sub-process, acceptance test level/sub-process |
| **test management** | planning, scheduling, estimating, monitoring, reporting, control and completion of test activities |
| **Test Management Process** | test process containing the sub-processes that are required for the management of a test project<br><br>Note 1 to entry: See Test Planning Process, Test Monitoring and Control Process, Test Completion Process. |
| **Test Monitoring and Control Process** | Test Management Process for ensuring that testing is performed in line with a Test Plan and with organizational test specifications |
| **Test Plan** | detailed description of test objectives to be achieved and the means and schedule for achieving them, organised to coordinate testing activities for some test item or set of test items |
| **Test Planning Process** | Test Management Process used to complete test planning and develop Test Plans |
| **test practice**<br>test approach | conceptual framework that can be applied to the Organizational Test Process, the Test Management Process, and/or the Dynamic Test Process to facilitate testing |
| **test procedure** | sequence of test cases in execution order, and any associated actions that may be required to set up the initial preconditions and any wrap up activities post execution |
| **Test Procedure Specification**<br>manual test script | document specifying one or more test procedures, which are collections of test cases to be executed for a particular objective |
| **test process** | provides information on the quality of a software product, often comprised of a number of activities, grouped into one or more test sub-processes |
| **test result** | indication of whether or not a specific test case has passed or failed, i.e. if the actual result observed as test item output corresponds to the expected result or if deviations were observed |
| **test script** | test procedure specification for manual or automated testing |
| **test set** | set of one or more test cases with a common constraint on their execution |
| **test specification** | complete documentation of the test design, test cases and test procedures for a specific test item |
| **test status report** | report that provides information about the status of the testing that is being performed in a specified reporting period |
| **test strategy** | part of the Test Plan that describes the approach to testing for a specific test project or test sub-process or sub-processes<br><br>Note 1 to entry: The test strategy is a distinct entity from the Organizational Test Strategy.<br>Note 2 to entry: The test strategy usually describes some or all of the following: the test practices used; the test sub- processes to be implemented; the retesting and regression testing to be employed; the test design techniques and corresponding test completion criteria to be used; test data; test environment and testing tool requirements; and expectations for test deliverables. |
| **test sub-process** | test management and dynamic (and static) test processes used to perform a specific test level (e.g. system testing, acceptance testing) or test type (e.g. usability testing, performance testing) normally within the context of an overall test process for a test project<br><br>Note 1 to entry: A test sub-process could comprise one or more test types. Depending on the life cycle model used, test sub-processes are also typically called test phases, test levels, test stages or test tasks. |
| **test traceability matrix**<br>verification cross reference matrix; requirements test matrix; requirements verification table | document, spreadsheet, or other automated tool used to identify related items in documentation and software, such as requirements with associated tests |
| **test type** | group of testing activities that are focused on specific quality characteristics<br><br>EXAMPLE Security testing, functional testing, usability testing, and performance testing. |
| **testing** | set of activities conducted to facilitate discovery and/or evaluation of properties of one or more test items |
| **testware** | artefacts produced during the test process required to plan, design, and execute tests<br><br>Note 1 to entry: Testware can include such things as documentation, scripts, inputs, expected results, files, databases, environment, and any additional software or utilities used in the course of testing. |
| **unscripted testing** | dynamic testing in which the tester's actions are not prescribed by written instructions in a test case |
| **volume testing** | type of performance efficiency testing conducted to evaluate the capability of the test item to process specified volumes of data (usually at or near maximum specified capacity) in terms of throughput capacity, storage capacity |

# Appendix III: TMMi glossary terms and their classification

*This Appendix contains all terms, synonyms and definitions of the TMMi software testing glossary. Also, we included the classification of all terms considering the 8 categories mentioned in Appendix I.*

| **TMMi Glossary v1.2** | **TMMi Terms Classified into 8 Categories or Conceptual Blocks for Testing.** Note that we include the synonyms of the terms in parentheses and not in bold. |
|---|---|
| **Cat#1 (C1): Test Project-, Strategy-, Organizational Test-related Terms** | |
| **exhaustive testing; risk-based testing; Test Process Improvement (TPI); test strategy; test process asset library; action proposal; Test Process Group (TPG); improvement proposal; independence of testing; requirements-based testing; resumption criteria; suspension criteria; test charter; test schedule; test infrastructure; test management; test phase; test policy; test approach** | |
| **Cat#2 (C2): Testing Work Process-, Activity-related Terms** | |
| **acceptance testing; beta testing; component integration testing; component testing** (unit testing); **formal review; informal review; inspection; peer review; incident logging; incident management; operational profile testing; re-testing** (confirmation testing); **review; static analysis; static testing; technical review; test closure; test control; test cycle; test design; test execution phase; test implementation; test level; test logging; test monitoring; test process; test run; test session; test type; testability review; testing; white-box testing; black-box testing; dynamic testing; alpha testing; functional testing; integration testing; non-functional testing; regression testing; system integration testing; system testing; test execution; test planning** | |
| **Cat#3 (C3): Test Goal-, Requirements-, Entity-related Terms** | |
| **Component; driver; stub; test condition; test harness; test environment; test item; test object; test objective** | |
| **Cat#4 (C4): Test Work Product-related Terms (e.g. Artifacts, Reports, Results, Specifications)** | |
| **branch coverage; checklist; condition determination coverage; decision coverage; defect density; Defect Detection Percentage (DDP); defect report; expected result; incident report; master test plan** (project test plan); **reliability growth model; result; smoke test; test data; test estimation; test execution schedule; test improvement plan; test input; test log; test procedure specification; test progress report; test script; test specification; test suite; test summary report; testware; intake test; incident; level test plan; phase test plan; release note; statement coverage; test; test basis; test case; test case specification; test design specification; test evaluation report; test plan; actual result; test performance indicator; condition coverage** | |
| **Cat#5 (C5): Testing Method-, Technique-related, Procedure-related, Rule-related Terms** | |
| **black-box test design technique; branch testing; cause-effect graphing; classification tree method; code coverage; decision table testing; elementary comparison testing; equivalence partitioning; error guessing; experienced-based test design technique; exploratory testing; heuristic evaluation; pass/fail criteria; state transition testing; statement testing; statistical testing; syntax testing; Test Point Analysis (TPA); use case testing; walkthrough; white-box test design technique; defect based test design technique; non-functional test design techniques; test design technique; boundary value analysis; condition testing; decision testing** | |
| **Cat#6 (C6): Testing Agent-, Role-, Tool-related Terms** | |
| **capture/playback tool; coverage tool; defect management tool; dynamic analysis tool; incident management tool; moderator; monitor; reviewer; review tool; Scribe; simulator; static code analyzer; test tool; tester; test comparator; test data preparation tool; test design tool; test execution tool; test manager; test management tool** | |
| **Cat#7 (C7): Other Terms somewhat related to Test (e.g., anomaly, defect, etc.)** | |
| **acceptance criteria; audit; defect; defect classification scheme; defect management; defect masking; defect taxonomy; equivalence partition; failure; Mean Time Between Failures (MTBF); product risk; project risk; risk; risk level; risk type; common causes; root cause; confidence level; entry criteria; exit criteria; horizontal traceability; Requirements management tool; input; post condition; precondition** | |
| **Cat#8 (C8): Terms beyond to the Test Domain related to Quality and Software Engineering** | |

**availability; best practice; Capability Maturity Model Integration (CMMI); configuration; configuration auditing; configuration control; configuration control board (CCB); Configuration identification; configuration item; configuration management; Configuration management tool; continuous representation; debugging tool; defect prevention; defined process; deliverable; emulator; error; feature; Function Point Analysis (FPA); functionality; generic goal; generic practice; Goal Question Metric (GQM); higher level management; indicator; institutionalization; integration; managed process; management review; maturity level; measure; measured process; measurement; measurement scale; metric; milestone; operational profile; optimizing process; output; Pareto analysis; priority; process; process area; process assessment; process capability; process improvement; process performance; process performance baseline; process performance objectives; project; quality assurance; quality attribute; quantitatively managed process; requirement; requirements phase; risk analysis; risk control** (risk mitigation); **risk identification**; **risk management; root cause analysis; rule; sampling; severity; software lifecycle; specific goal; specific practice; specification; specified input; staged representation; standard; statistical process control; statistical technique; statistically managed process; status accounting; sub-practice; system; Test Maturity Model (TMM); Test Maturity Model integration (TMMi); traceability; trustworthiness; usability; V-model; validation; verification; vertical traceability; Wide Band Delphi; impact analysis; Mean Time To Repair (MTTR); efficiency; maintainability; performance indicator; portability; reliability**

| **Term name**<br>synonyms | **Definition** |
|---|---|
| **acceptance criteria** | The exit criteria that a component or system must satisfy in order to be accepted by a user, customer, or other authorized entity. [IEEE 610] |
| **acceptance testing** | Formal testing with respect to user needs, requirements, and business processes conducted to determine whether or not a system satisfies the acceptance criteria and to enable the user, customers or other authorized entity to determine whether or not to accept the system. [After IEEE 610] |
| **action proposal** | The documented action to be taken to prevent the future occurrence of common causes of defects or to incorporate best practices into test process assets. |
| **actual result** | The behavior produced/observed when a component or system is tested. |
| **alpha testing** | Simulated or actual operational testing by potential users/customers or an independent test team at the developers' site, but outside the development organization. Alpha testing is often employed for off-the-shelf software as a form of internal acceptance testing. |
| **audit** | An independent evaluation of software products or processes to ascertain compliance to standards, guidelines, specifications, and/or procedures based on objective criteria, including documents that specify:<br>(1) the form or content of the products to be produced<br>(2) the process by which the products shall be produced<br>(3) how compliance to standards or guidelines shall be measured. [IEEE 1028] |
| **availability** | The degree to which a system or component is operational and accesible when required to use. [IEEE 610] |
| **best practice** | A superior method or innovative practice that contributes to the improved performance of an organization within a given context, usually recognized as 'best' by other peer organizations. |
| **beta testing** | Operational testing by potential and/or existing users/customers at an external site not otherwise involved with the developers, to determine whether or not a component or system satisfies the user/customer needs and fits within the business processes. Beta testing is often employed as a form of external acceptance testing for off-the-shelf software in order to acquire feedback from the market. |
| **black-box testing** | Testing, either functional or non-functional, without reference to the internal structure of the component or system. |
| **black-box test design technique** | Technique/procedure to derive and/or select test cases based on an analysis of the specification, either functional or non-functional, of a component or system without reference to its internal structure. |
| **boundary value analysis** | A black box test design technique in which test cases are designed based on boundary values. |
| **branch coverage** | The percentage of branches that have been exercised by a test suite. 100% branch coverage implies both 100% decision coverage and 100% statement coverage. |
| **branch testing** | A white box test design technique in which test cases are designed to execute branches. |
| **Capability Maturity Model Integration (CMMI)** | A framework that describes the key elements of an effective product development and maintenance process. The Capability Maturity Model Integration covers best practices for planning, engineering and managing product development and maintenance. [CMMI] |
| **capture/playback tool** | A type of test execution tool where inputs are recorded during manual testing in order to generate automated test scripts that can be executed later (i.e. replayed). These tools are often used to support automated regression testing. |
| **cause-effect graphing** | A black box test design technique in which test cases are designed from cause-effect graphs. [BS 7925/2] |
| **classification tree method** | A black box test design technique in which test cases, described by means of a classification tree, are designed to execute combinations of representatives of input and/or output domains. [Grochtmann] |
| **checklist** | Checklists are 'stored wisdom' aimed at helping to interpret the rules and explain their application. Checklists are used to increase effectiveness at finding major defects in a specification during a review. A checklist usually takes the form of a list of questions. All checklist questions are derived directly and explicitly from cross-referenced specification rules. [Gilb and Graham] |
| **code coverage** | An analysis method that determines which parts of the software have been executed (covered) by the test suite and which parts have not been executed, e.g., statement coverage, decision coverage or condition coverage. |
| **common causes** | The underlying source of a number of defects of a similar type, so that if the root cause is addressed the occurrence of these types of defects is decreased or removed. |
| **component** | A minimal software item that can be tested in isolation. |
| **component integration testing** | Testing performed to expose defects in the interfaces and interaction between integrated components. |
| **component testing** unit testing | The testing of individual software components. [After IEEE 610] |
| **condition coverage** | The percentage of condition outcomes that have been exercised by a test suite. 100% condition coverage requires each single condition in every decision statement to be tested as True and False. |
| **condition determination coverage** | The percentage of all single condition outcomes that independently affect a decision outcome that have been exercised by a test case suite. 100% modified condition decision coverage implies 100% decision condition coverage. |
| **condition testing** | A white box test design technique in which test cases are designed to execute condition outcomes. |
| **confidence level** | The likelihood that the software is defect-free. [Burnstein] |
| **configuration** | The composition of a component or system as defined by the number, nature, and interconnections of its constituent parts. |
| **configuration auditing** | The function to check on the contents of libraries of configuration items, e.g., for standards compliance. [IEEE 610] |
| **configuration control** | An element of configuration management, consisting of the evaluation, coordination, approval or disapproval, and implementation of changes to configuration items after formal establishment of their configuration identification. [IEEE 610] |
| **configuration control board (CCB)** | A group of people responsible for evaluating and approving or disapproving proposed changes to configuration items, and for ensuring implementation of approved changes. [IEEE 610] |
| **Configuration identification** | An element of configuration management, consisting of selecting the configuration items for a system and recording their functional and physical characteristics in technical documentation. [IEEE 610] |
| **configuration item** | An aggregation of hardware, software or both, that is designated for configuration management and treated as a single entity in the configuration management process. [IEEE 610] |
| **configuration management** | A discipline applying technical and administrative direction and surveillance to: identify and document the functional and physical characteristics of a configuration item, control changes to those characteristics, record and report change processing and implementation status, and verify compliance with specified requirements. [IEEE 610] |
| **Configuration management tool** | A tool that provides support for the identification and control of configuration items, their status over changes and versions, and the release of baselines consisting of configuration items. |
| **continuous representation** | A capability maturity model structure wherein capability levels provide a recommended order for approaching process improvement within specified process areas. [CMMI] |
| **coverage tool** | A tool that provides objective measures of what structural elements, e.g., statements and/or branches, have been exercised by a test suite. |
| **debugging tool** | A tool used by programmers to reproduce failures, investigate the state of programs and find the corresponding defect. Debuggers enable programmers to execute programs step by step, to halt a program at any program statement and to set and examine program variables. |
| **decision coverage** | The percentage of decision outcomes that have been exercised by a test suite. 100% decision coverage implies both 100% branch coverage and 100% statement coverage. |
| **decision table testing** | A black box test design technique in which test cases are designed to execute the combinations of inputs and/or stimuli (causes) shown in a decision table. [Veenendaal] |
| **decision testing** | A white box test design technique in which test cases are designed to execute decision outcomes. |
| **defect** | A flaw in a component or system that can cause the component or system to fail to perform its required function, e.g., an incorrect statement or data definition. A defect, if encountered during execution, may cause a failure of the component or system. |
| **defect based test design technique** | A procedure to derive and/or select test cases targeted at one or more defect categories, with tests being developed from what is known about the specific defect category. See also defect taxonomy. |
| **defect classification scheme** | A set of categories, including phase, defect type, cause, severity, priority, to describe a defect in a consistent manner. |
| **defect density** | The number of defects identified in a component or system divided by the size of the component or system (expressed in standard measurement terms, e.g., lines-of-code, number of classes or function points). |
| **Defect Detection Percentage (DDP)** | The number of defects found by a test phase, divided by the number found by that test phase and any other means afterwards. |
| **defect management** | The process of recognizing, investigating, taking action and disposing of defects. It involves recording defects, classifying them and identifying the impact. [After IEEE 1044] |
| **defect management tool** | A tool that facilitates the recording and status tracking of defects and changes. They often have workflow-oriented facilities to track and control the allocation, correction and re-testing of defects and provide reporting facilities. See also incident management tool. |
| **defect masking** | An occurrence in which one defect prevents the detection of another. [After IEEE 610] |
| **defect prevention** | The activities involved in identifying defects or potential defects, analyzing these defects to find their root causes and preventing them from being introduced into future products. [After Burnstein] |
| **defect report** | A document reporting on any flaw in a component or system that can cause the component or system to fail to perform its required function. [After IEEE 829] |
| **defect taxonomy** | A system of (hierarchical) categories designed to be a useful aid for reproducibly classifying defects. |
| **defined process** | A managed process that is tailored from the organization's set of standard processes according to the organization's tailoring guidelines; has maintained process description; and contributes work products, measures, and other process improvement information to the organizational process assets. [CMMI] |
| **deliverable** | Any (work) product that must be delivered to someone other than the (work) product's author. |
| **driver** | A software component or test tool that replaces a component that takes care of the control and/or the calling of a component or system. [After TMap] |
| **dynamic analysis tool** | A tool that provides run-time information on the state of the software code. These tools are most commonly used to identify unassigned pointers, check pointer |
| | arithmetic and to monitor the allocation, use and de-allocation of memory and to flag memory leaks. |
| **dynamic testing** | Testing that involves the execution of the software of a component or system. |
| **efficiency** | The capability of the software product to provide appropriate performance relative to the amount of resources used under stated conditions. [ISO 9126] |
| **elementary comparison testing** | A black box test design technique in which test cases are designed to execute combinations of inputs using the concept of condition determination coverage. [TMap] |
| **emulator** | A device, computer program, or system that accepts the same inputs and produces the same outputs as a given system. [IEEE 610] See also simulator. |
| **entry criteria** | The set of generic and specific conditions for permitting a process to go forward with a defined task, e.g., test phase. The purpose of entry criteria is to prevent a task from starting which would entail more (wasted) effort compared to the effort needed to remove the failed entry criteria. [Gilb and Graham] |
| **equivalence partition** | A portion of an input or output domain for which the behavior of a component or system is assumed to be the same, based on the specification. |
| **equivalence partitioning** | A black box test design technique in which test cases are designed to execute representatives from equivalence partitions. In principle test cases are designed to cover each partition at least once. |
| **error** | A human action that produces an incorrect result. [After IEEE 610] |
| **error guessing** | A test design technique where the experience of the tester is used to anticipate what defects might be present in the component or system under test as a result of errors made, and to design tests specifically to expose them. |
| **exhaustive testing** | A test approach in which the test suite comprises all combinations of input values and preconditions. |
| **exit criteria** | The set of generic and specific conditions, agreed upon with the stakeholders, for permitting a process to be officially completed. The purpose of exit criteria is to prevent a task from being considered completed when there are still outstanding parts of the task which have not been finished. Exit criteria are used to report against and to plan when to stop testing. [After Gilb and Graham] |
| **expected result** | The behavior predicted by the specification, or another source, of the component or system under specified conditions. |
| **experienced-based test design technique** | Procedure to derive and/or select test cases based on the tester's experience, knowledge and intuition. |
| **exploratory testing** | An informal test design technique where the tester actively controls the design of the tests as those tests are performed and uses information gained while testing to design new and better tests. [After Bach] |
| **failure** | Deviation of the component or system from its expected delivery, service or result. [After Fenton] |
| **feature** | An attribute of a component or system specified or implied by requirements documentation (for example reliability, usability or design constraints). [After IEEE 1008] |
| **formal review** | A review characterized by documented procedures and requirements, e.g., inspection. |
| **Function Point Analysis (FPA)** | Method aiming to measure the size of the functionality of an information system. The measurement is independent of the technology. This measurement may be used as a basis for the measurement of productivity, the estimation of the needed resources, and project control. |
| **functional testing** | Testing based on an analysis of the specification of the functionality of a component or system. See also black box testing. |
| **functionality** | The capability of the software product to provide functions which meet stated and implied needs when the software is used under specified conditions. [ISO 9126] |
| **generic goal** | A required model component that describes the characteristics that must be present to institutionalize the processes that implement a process area. [CMMI] |
| **generic practice** | An expected model component that is considered important in achieving the associated generic goal. The generic practices associated with a generic goal describe the activities that are expected to result in achievement of the generic goal and contribute to the institutionalization of the processes associated with a process area. [CMMI] |
| **Goal Question Metric (GQM)** | An approach to software measurement using a three-level model: conceptual level (goal), operational level (question) and quantitative level (metric). |
| **heuristic evaluation** | A static usability test technique to determine the compliance of a user interface with recognized usability principles (the so-called "heuristics"). |
| **higher level management** | The person or persons who provide the policy and overall guidance for the process, but do not to provide direct day-to-day monitoring and controlling of the process. Such persons belong to a level of management in the organization above the intermediate level responsible for the process and can be (but are not necessarily) senior managers. [CMMI] |
| **horizontal traceability** | The tracing of requirements for a test level through the layers of test documentation (e.g., test plan, test design specification, test case specification and test procedure specification or test script). The horizontal traceability is expected to be bi-directional. |
| **impact analysis** | The assessment of change to the layers of development documentation, test documentation and components, in order to implement a given change to specified requirements. |
| **improvement proposal** | A change request that addresses a proposed process or technology improvement and typically also includes a problem statement, a plan for implementing the improvement, and quantitative success criteria for evaluating actual results of the deployment within the change process managed by the Test Process Group. |
| **Incident** | Any event occurring that requires investigation. [After IEEE 1008] |
| **incident logging** | Recording the details of any incident that occurred, e.g., during testing. |
| **incident management** | The process of recognizing, investigating, taking action and disposing of incidents. It includes logging incidents, classifying them and identifying the impact. [After IEEE 1044] |
| **incident management tool** | A tool that facilitates the recording and status tracking of incidents. They often have workflow-oriented facilities to track and control the allocation, correction and re-testing of incidents and provide reporting facilities. See also defect management tool. |
| **incident report** | A document reporting on any event that occurred, e.g., during the testing, which requires investigation. [After IEEE 829] |
| **independence of testing** | Separation of responsibilities, which encourages the accomplishment of objective testing. [After DO-178b] |
| **indicator** | A measure that can be used to estimate or predict another measure. [ISO 14598] |
| **informal review** | A review not based on a formal (documented) procedure. |
| **input** | A variable (whether stored within a component or outside) that is read by a component. |
| **inspection** | A type of peer review that relies on visual examination of documents to detect defects, e.g., violations of development standards and non-conformance to higher level documentation. The most formal review technique and therefore always based on a documented procedure. [After IEEE 610, IEEE 1028] See also peer review. |
| **institutionalization** | The ingrained way of doing business that an organization follows routinely as part of its corporate culture. |
| **intake test** | A special instance of a smoke test to decide if the component or system is ready for detailed and further testing. An intake test is typically carried out at the start of the test execution phase. See also smoke test. |
| **integration** | The process of combining components or systems into larger assemblies. |
| **integration testing** | Testing performed to expose defects in the interfaces and in the interactions between integrated components or systems. See also component integration testing, system integration testing. |
| **level test plan** | A test plan that typically addresses one test level. See also test plan. |
| **maintainability** | The ease with which a software product can be modified to correct defects, modified to meet new requirements, modified to make future maintenance easier, or adapted to a changed environment. [ISO 9126] |
| **managed process** | A performed process that is planned and executed in accordance with policy; employs skilled people having adequate resources to produce controlled outputs; involves relevant stakeholders; is monitored, controlled and reviewed; and is evaluated for adherence to its process description. [CMMI] |
| **management review** | A systematic evaluation of software acquisition, supply, development, operation, or maintenance process, performed by or on behalf of management that monitors |
| | progress, determines the status of plans and schedules, confirms requirements and their system allocation, or evaluates the effectiveness of management approaches to achieve fitness for purpose. [After IEEE 610, IEEE 1028] |
| **master test plan**<br>project test plan | A test plan that typically addresses multiple test levels. See also test plan. |
| **maturity level** | Degree of process improvement across a predefined set of process areas in which all goals in the set are attained. [CMMI] |
| **Mean Time Between Failures (MTBF)** | The arithmetic mean (average) time between failures of a system. The MTBF is typically part of a reliability growth model that assumes the failed system is immediately repaired, as a part of a defect fixing process. |
| **Mean Time To Repair (MTTR)** | The arithmetic mean (average) time a system will take to recover from any failure. This typically includes testing to insure that the defect has been resolved. |
| **measure** | The number or category assigned to an attribute of an entity by making a measurement. [ISO 14598] |
| **measured process** | A defined process whereby product quality and process attributes are consistently measured, and the measures are used to improve and make decisions regarding product quality and process-performance. |
| **measurement** | The process of assigning a number or category to an entity to describe an attribute of that entity. [ISO 14598] |
| **measurement scale** | A scale that constrains the type of data analysis that can be performed on it. [ISO 14598] |
| **metric** | A measurement scale and the method used for measurement. [ISO 14598] |
| **milestone** | A point in time in a project at which defined (intermediate) deliverables and results should be ready. |
| **moderator** | The leader and main person responsible for an inspection or other review process. |
| **monitor** | A software tool or hardware device that runs concurrently with the component or system under test and supervises, records and/or analyzes the behavior of the component or system. [After IEEE 610] |
| **non-functional testing** | Testing the attributes of a component or system that do not relate to functionality, e.g., reliability, efficiency, usability, maintainability and portability. |
| **non-functional test design techniques** | Procedure to derive and/or select test cases for non-functional testing based on an analysis of the specification of a component or system without reference to its internal structure. See also black box test design technique. |
| **operational profile** | The representation of a distinct set of tasks performed by the component or system, possibly based on user behavior when interacting with the component or system, and their probabilities of occurrence. A task is logical rather tha physical and can be executed over several machines or be executed in non-contiguous time segments. |
| **operational profile testing** | Statistical testing using a model of system operations (short duration tasks) and their probability of typical use. [Musa] |
| **optimizing process** | A quantitatively managed process that is improved based on an understanding of the common causes of variation inherent in the process. The focus of an optimizing process is on continually improving the range of process performance through both incremental and innovative improvements. |
| **output** | A variable (whether stored within a component or outside) that is written by a component. |
| **Pareto analysis** | A statistical technique in decision making that is used for selection of a limited number of factors that produce significant overall effect. In terms of quality improvement, a large majority of problems (80%) are produced by a few key causes (20%). |
| **pass/fail criteria** | Decision rules used to determine whether a test item (function) or feature has passed or failed a test. [IEEE 829] |
| **peer review** | A review of a software work product by colleagues of the producer of the product for the purpose of identifying defects and improvements. Examples are inspection, technical review and walkthrough. |
| **performance indicator** | A high level metric of effectiveness and/or efficiency used to guide and control progressive development, e.g., lead-time slip for software development. [CMMI] |
| **phase test plan** | A test plan that typically addresses one test phase. See also test plan. |
| **portability** | The ease with which the software product can be transferred from one hardware or software environment to another. [ISO 9126] |
| **post condition** | Environmental and state conditions that must be fulfilled after the execution of a test or test procedure. |
| **precondition** | Environmental and state conditions that must be fulfilled before the component or system can be executed with a particular test or test procedure. |
| **priority** | The level of (business) importance assigned to an item, e.g., defect. |
| **process** | A set of interrelated activities, which transform inputs into outputs. [ISO 12207] |
| **process area** | A cluster of related practices in an area that, when implemented collectively, satisfy a set of goals considered important for making improvements in that area. [CMMI] |
| **process assessment** | A disciplined evaluation of an organization's software processes against a reference model. [after ISO 15504] |
| **process capability** | The range of expected results that can be achieved by following a process. |
| **process improvement** | A program of activities designed to improve the performance and maturity of the organization's processes, and the result of such a program. [CMMI] |
| **process performance** | A measure of actual results achieved by following a process. [CMMI] |
| **process performance baseline** | A documented characterization of the actual results achieved by following a process, which is used as a benchmark for comparing actual process performance against expected process performance. [CMMI] |
| **process performance objectives** | Objectives and requirements for product quality, service quality and process performance. |
| **product risk** | A risk directly related to the test object. See also risk. |
| **project** | A project is a unique set of coordinated and controlled activities with start and finish dates undertaken to achieve an objective conforming to specific requirements, including the constraints of time, cost and resources. [ISO 9000] |
| **project risk** | A risk related to management and control of the (test) project, e.g., lack of staffing, strict deadlines, changing requirements, etc. See also risk. |
| **quality assurance** | Part of quality management focused on providing confidence that quality requirements will be fulfilled. [ISO 9000] |
| **quality attribute** | A feature or characteristic that affects an item's quality. [IEEE 610] |
| **quantitatively managed process** | A defined process that is controlled using statistical and other quantitative techniques. The product quality, service quality, and process-performance attributes are measured and controlled throughout the project. [CMMI] |
| **regression testing** | Testing of a previously tested program following modification to ensure that defects have not been introduced or uncovered in unchanged areas of the software, as a result of the changes made. It is performed when the software or its environment is changed. |
| **release note** | A document identifying test items, their configuration, current status and other delivery information delivered by development to testing, and possibly other stakeholders, at the start of a test execution phase. [After IEEE 829] |
| **reliability** | The capability of the software product to perform its required functions under stated conditions for a specified period of time, or for a specified number of operations. [ISO 9126] |
| **reliability growth model** | A model that shows the growth in reliability over time during continuous testing of a component or system as a result of the removal of defects that result in reliability failures. |
| **requirement** | A condition or capability needed by a user to solve a problem or achieve an objective that must be met or possessed by a system or system component to satisfy a contract, standard, specification, or other formally imposed document. [After IEEE 610] |
| **requirements-based testing** | An approach to testing in which test cases are designed based on test objectives and test conditions derived from requirements, e.g., tests that exercise specific functions or probe non-functional attributes such as reliability or usability. |
| **Requirements management tool** | A tool that supports the recording of requirements, requirements attributes (e.g., priority, knowledge responsible) and annotation, facilitates traceability through layers of requirements and to test cases, and requirements change management. Some requirements management tools also provide facilities for static analysis, such as consistency checking and violations to pre-defined requirements rules. |
| **requirements phase** | The period of time in the software lifecycle during which the requirements for a software product are defined and documented. [IEEE 610] |
| **result** | The consequence/outcome of the execution of a test. It includes outputs to screens, changes to data, reports, and communication messages sent out. See also actual result, expected result. |
| **resumption criteria** | The testing activities that must be repeated when testing is re-started after a suspension. [After IEEE 829] |
| **re-testing**<br>confirmation testing | Testing that runs test cases that failed the last time they were run, in order to verify the success of corrective actions. |
| **review** | An evaluation of a product or project status to ascertain discrepancies from planned results and to recommend improvements. Examples include management review, informal review, technical review, inspection, and walkthrough. [After IEEE 1028] |
| **reviewer** | The person involved in the review that identifies and describes anomalies in the product or project under review. Reviewers can be chosen to represent different viewpoints and roles in the review process. |
| **review tool** | A tool that provides support to the review process. Typical features include review planning and tracking support, communication support, collaborative reviews and a repository for collecting and reporting of metrics. |
| **risk** | A factor that could result in future negative consequences; usually expressed as impact and likelihood. |
| **risk analysis** | The process of assessing identified risks to estimate their impact and probability of occurrence (likelihood). |
| **risk-based testing** | An approach to testing to reduce the level of product risks and inform stakeholders on their status, starting in the initial stages of a project. It involves the identification of product risks and their use in guiding the test process. |
| **risk control**<br>risk mitigation | The process through which decisions are reached and protective measures are implemented for reducing risks to, or maintaining risks within, specified levels. |
| **risk identification** | The process of identifying risks using techniques such as brainstorming, checklists and failure history. |
| **risk level** | The importance of a risk as defined by its characteristics, impact and likelihood. The level of risk can be used to determine the intensity of testing to be performed. A risk level can be expressed either qualitatively (e.g., high, medium, low) or quantitatively. |
| **risk management** | Systematic application of procedures and practices to the tasks of identifying, analyzing, prioritizing, and controlling risk. |
| **risk type** | A specific category of risk related to the type of testing that can mitigate (control) that category. For example the risk of user-interactions being misunderstood can be mitigated by usability testing. |
| **root cause** | A source of a defect such that if it is removed, the occurrence of the defect type is decreased or removed. [CMMI] |
| **root cause analysis** | An analysis technique aimed at identifying the root causes of defects. By directing corrective measures at root causes, it is hoped that the likelihood of defect recurrence will be minimized. |
| **rule** | A rule is any statement of a standard on how to write or carry out some part of a systems engineering or business process. [Gilb and Graham] |
| **sampling** | A statistical practice concerned with the selection of an unbiased or random subset of individual observations within a population of individuals intended to yield some knowledge about the population of concern as a whole. |
| **scribe** | The person who records each defect mentioned and any suggestions for process improvement during a review meeting, on a logging form. The scribe has to ensure that the logging form is readable and understandable. |
| **severity** | The degree of impact that a defect has on the development or operation of a component or system. [After IEEE 610] |
| **simulator** | A device, computer program or system used during testing, which behaves or operates like a given system when provided with a set of controlled inputs. [After IEEE 610, DO178b] See also emulator. |
| **smoke test** | A subset of all defined/planned test cases that cover the main functionality of a component or system, ascertaining that the most crucial functions of a program work, but not bothering with finer details. A daily build and smoke test is among industry best practices. See also intake test. |
| **software lifecycle** | The period of time that begins when a software product is conceived and ends when the software is no longer available for use. The software lifecycle typically includes a |
| | concept phase, requirements phase, design phase, implementation phase, test phase, installation and checkout phase, operation and maintenance phase, and sometimes, retirement phase. Note these phases may overlap or be performed iteratively. |
| **specific goal** | A required model component that describes the unique characteristics that must be present to satisfy the process area. [CMMI] |
| **specific practice** | An expected model component that is considered important in achieving the associated specific goal. The specific practices describe the activities expected to result in achievement of the specific goals of a process area. [CMMI] |
| **specification** | A document that specifies, ideally in a complete, precise and verifiable manner, the requirements, design, behavior, or other characteristics of a component or system, and, often, the procedures for determining whether these provisions have been satisfied. [After IEEE 610] |
| **specified input** | An input for which the specification predicts a result. |
| **staged representation** | A model structure wherein attaining the goals of a set of process areas establishes a maturity level; each level builds a foundation for subsequent levels. [CMMI] |
| **standard** | Formal, possibly mandatory, set of requirements developed and used to prescribe consistent approaches to the way of working or to provide guidelines (e.g., ISO/IEC standards, IEEE standards, and organizational standards). [After CMMI] |
| **state transition testing** | A black box test design technique in which test cases are designed to execute valid and invalid state transitions. |
| **statement coverage** | The percentage of executable statements that have been exercised by a test suite. |
| **statement testing** | A white box test design technique in which test cases are designed to execute statements. |
| **static analysis** | Analysis of software artifacts, e.g., requirements or code, carried out without execution of these software artifacts. |
| **static code analyzer** | A tool that carries out static code analysis. The tool checks source code, for certain properties such as conformance to coding standards, quality metrics or data flow anomalies. |
| **static testing** | Testing of a component or system at specification or implementation level without execution of that software, e.g., reviews or static code analysis. |
| **statistical process control** | Statistically based analysis of a process and measurements of process performance, which will identify common and special causes of variation in the process performance, and maintain process performance within limits. [CMMI] |
| **statistical technique** | An analytical technique that employs statistical methods (e.g., statistical process control, confidence intervals, and prediction intervals). [CMMI] |
| **statistical testing** | A test design technique in which a model of the statistical distribution of the input is used to construct representative test cases. |
| **statistically managed process** | A process that is managed by a statistically based technique in which processes are analyzed, special causes of process variation are identified, and process performance is contained within well-defined limits. [CMMI] |
| **status accounting** | An element of configuration management, consisting of the recording and reporting of information needed to manage a configuration effectively. This information includes a listing of the approved configuration identification, the status of proposed changes to the configuration, and the implementation status of the approved changes. [IEEE 610] |
| **stub** | A skeletal or special-purpose implementation of a software component, used to develop or test a component that calls or is otherwise dependent on it. It replaces a called component. [After IEEE 610] |
| **sub-practice** | An informative model component that provides guidance for interpreting and implementing a specific or generic practice. Sub-practices may be worded as if prescriptive, but are actually meant only to provide ideas that may be useful for process improvement. [CMMI] |
| **suspension criteria** | The criteria used to (temporarily) stop all or a portion of the testing activities on the test items. [After IEEE 829] |
| **syntax testing** | A black box test design technique in which test cases are designed based upon the definition of the input domain and/or output domain. |
| **system** | A collection of components organized to accomplish a specific function or set of functions. [IEEE 610] |
| **system integration testing** | Testing the integration of systems and packages; testing interfaces to external organizations (e.g., Electronic Data Interchange, Internet). |
| **system testing** | The process of testing an integrated system to verify that it meets specified requirements. [Hetzel] |
| **technical review** | A peer group discussion activity that focuses on achieving consensus on the technical approach to be taken. [Gilb and Graham, IEEE 1028] See also peer review. |
| **test** | A set of one or more test cases. [IEEE 829] |
| **test approach** | The implementation of the test strategy for a specific project. It typically includes the decisions made that consider the (test) project's goal and the risk assessment carried out, starting points regarding the test process, the test design techniques to be applied, exit criteria and test types to be performed. |
| **test basis** | All documents from which the requirements of a component or system can be inferred. The documentation on which the test cases are based. If a document can be amended only by way of a formal amendment procedure, then the test basis is called a frozen test basis. [After TMap] |
| **test case** | A set of input values, execution preconditions, expected results and execution post conditions, developed for a particular objective or test condition, such as to exercise a particular program path or to verify compliance with a specific requirement. [After IEEE 610] |
| **test case specification** | A document specifying a set of test cases (objective, inputs, test actions, expected results, and execution preconditions) for a test item. [After IEEE 829] |
| **test charter** | A statement of test objectives, and possibly test ideas about how to test. Test charters are used in exploratory testing. See also exploratory testing. |
| **test closure** | During the test closure phase of a test process data is collected from completed activities to consolidate experience, test ware, facts and numbers. The test closure phase consists of finalizing and archiving the test ware and evaluating the test process, including preparation of a test evaluation report. See also test process. |
| **test comparator** | A test tool to perform automated test comparison of actual results with expected results. |
| **test condition** | An item or event of a component or system that could be verified by one or more test cases, e.g., a function, transaction, feature, quality attribute, or structural element. |
| **test control** | A test management task that deals with developing and applying a set of corrective actions to get a test project on track when monitoring shows a deviation from what was planned. See also test management. |
| **test cycle** | Execution of the test process against a single identifiable release of the test object. |
| **test data** | Data that exists (for example, in a database) before a test is executed, and that affects or is affected by the component or system under test. |
| **test data preparation tool** | A type of test tool that enables data to be selected from existing databases or created, generated, manipulated and edited for use in testing. |
| **test design** | (1) See test design specification.<br>(2) The process of transforming general testing objectives into tangible test conditions and test cases. |
| **test design specification** | A document specifying the test conditions (coverage items) for a test item, the detailed test approach and identifying the associated high level test cases. [After IEEE 829] |
| **test design technique** | Procedure used to derive and/or select test cases. |
| **test design tool** | A tool that supports the test design activity by generating test inputs from a specification that may be held in a CASE tool repository, e.g., requirements management tool, from specified test conditions held in the tool itself, or from code. |
| **test environment** | An environment containing hardware, instrumentation, simulators, software tools, and other support elements needed to conduct a test. [After IEEE 610] |
| **test estimation** | The calculated approximation of a result (e.g., effort spent, completion date, costs involved, number of test cases, etc.) which is usable even if input data may be incomplete, uncertain, or noisy. |
| **test evaluation report** | A document produced at the end of the test process summarizing all testing activities and results. It also contains an evaluation of the test process and lessons learned. |
| **test execution** | The process of running a test on the component or system under test, producing actual result(s). |
| **test execution phase** | The period of time in a software development lifecycle during which the components of a software product are executed, and the software product is evaluated to determine whether or not requirements have been satisfied. [IEEE 610] |
| **test execution schedule** | A scheme for the execution of test procedures. The test procedures are included in the test execution schedule in their context and in the order in which they are to be executed. |
| **test execution tool** | A type of test tool that is able to execute other software using an automated test script, e.g., capture/playback. [Fewster and Graham] |
| **test harness** | A test environment comprised of stubs and drivers needed to execute a test. |
| **test implementation** | The process of developing and prioritizing test procedures, creating test data and, optionally, preparing test harnesses and writing automated test scripts. |
| **test improvement plan** | A plan for achieving organizational test process improvement objectives based on a thorough understanding of the current strengths and weaknesses of the organization's test processes and test process assets. [After CMMI] |
| **test infrastructure** | The organizational artifacts needed to perform testing, consisting of test environments, test tools, office environment and procedures. |
| **test input** | The data received from an external source by the test object during test execution. The external source can be hardware, software or human. |
| **test item** | The individual element to be tested. There usually is one test object and many test items. See also test object. |
| **test level** | A group of test activities that are organized and managed together. A test level is linked to the responsibilities in a project. Examples of test levels are component test, integration test, system test and acceptance test. [After TMap] |
| **test log** | A chronological record of relevant details about the execution of tests. [IEEE 829] |
| **test logging** | The process of recording information about tests executed into a test log. |
| **test manager** | The person responsible for project management of testing activities and resources, and evaluation of a test object. The individual who directs, controls, administers, plans and regulates the evaluation of a test object. |
| **test management** | The planning, estimating, monitoring and control of test activities, typically carried out by a test manager. |
| **test management tool** | A tool that provides support to the test management and control part of a test process. It often has several capabilities, such as test ware management, scheduling of tests, the logging of results, progress tracking, incident management and test reporting. |
| **Test Maturity Model (TMM)** | A five level staged framework for test process improvement, related to the Capability Maturity Model (CMM), which describes the key elements of an effective test process. |
| **Test Maturity Model integration (TMMi)** | A five level staged framework for test process improvement, related to the Capability Maturity Model Integration (CMMI), which describes the key elements of an effective test process. |
| **test monitoring** | A test management task that deals with the activities related to periodically checking the status of a test project. Reports are prepared that compare the actuals to that which was planned. See also test management. |
| **test object** | The component or system to be tested. See also test item. |
| **test objective** | A reason or purpose for designing and executing a test. |
| **test performance indicator** | A high level metric of effectiveness and/or efficiency used to guide and control progressive test development, e.g., Defect Detection Percentage (DDP). |
| **test phase** | A distinct set of test activities collected into a manageable phase of a project, e.g., the execution activities of a test level. [After Gerrard] |
| **test plan** | A document describing the scope, approach, resources and schedule of intended test activities. It identifies amongst others test items, the features to be tested, the testing tasks, who will do each task, degree of tester independence, the test environment, the test design techniques and entry and exit criteria to be used, and the rationale for their choice, and any risks requiring contingency planning. It is a record of the test planning process. [After IEEE 829] |
| **test planning** | The activity of establishing or updating a test plan. |
| **test policy** | A high level document describing the principles, approach and major objectives of the organization regarding testing. |
| **Test Point Analysis (TPA)** | A formula based test estimation method based on function point analysis. [TMap] |
| **test procedure specification** | A document specifying a sequence of actions for the execution of a test. Also known as test script or manual test script. [After IEEE 829] |
| **test process** | The fundamental test process comprises test planning and control, test analysis and design, test implementation and execution, evaluating exit criteria and reporting, and test closure activities. |
| **test process asset library** | A collection of test process asset holdings that can be used by an organization or project. [CMMI] |
| **Test Process Group (TPG)** | A permanent or virtual entity in the organization responsible for test process related activities such as process definition, analysis and assessment, action planning and evaluation. It has the overall test process ownership as defined in an organization's test policy. |
| **Test Process Improvement (TPI)** | A continuous framework for test process improvement that describes the key elements of an effective test process, especially targeted at system testing and acceptance testing. |
| **test progress report** | A document summarizing testing activities and results, produced at regular intervals, to report progress of testing activities against a baseline (such as the original test plan) and to communicate risks and alternatives requiring a decision to management. |
| **test run** | Execution of a test on a specific version of the test object. |
| **test schedule** | A list of activities, tasks or events of the test process, identifying their intended start and finish dates and/or times, and interdependencies. |
| **test script** | Commonly used to refer to a test procedure specification, especially an automated one. |
| **test session** | An uninterrupted period of time spent in executing tests. In exploratory testing, each test session is focused on a charter, but testers can also explore new opportunities or issues during a session. The tester creates and executes test cases on the fly and records their progress. See also exploratory testing. |
| **test specification** | A document that consists of a test design specification, test case specification and/or test procedure specification. |
| **test strategy** | A high-level description of the test levels to be performed and the testing within those levels for an organization or programme (one or more projects). |
| **test suite** | A set of several test cases for a component or system under test, where the post condition of one test is often used as the precondition for the next one. |
| **test summary report** | A document summarizing testing activities and results. It also contains an evaluation of the corresponding test items against exit criteria. [After IEEE 829] |
| **test tool** | A software product that supports one or more test activities, such as planning and control, specification, building initial files and data, test execution and test analysis. [TMap] |
| **test type** | A group of test activities aimed at testing a component or system focused on a specific test objective, i.e. functional test, usability test, regression test etc. A test type may take place on one or more test levels or test phases. [After TMap] |
| **testability review** | A detailed check of the test basis to determine whether the test basis is at an adequate quality level to act as an input document for the test process. [After TMap] |
| **tester** | A skilled professional who is involved in the testing of a component or system. |
| **testing** | The process consisting of all lifecycle activities, both static and dynamic, concerned with planning, preparation and evaluation of software products and related work products to determine that they satisfy specified requirements, to demonstrate that they are fit for purpose and to detect defects. |
| **testware** | Artifacts produced during the test process required to plan, design, and execute tests, such as documentation, scripts, inputs, expected results, set-up and clear-up procedures, files, databases, environment, and any additional software or utilities used in testing. [After Fewster and Graham] |
| **traceability** | The ability to identify related items in documentation and software, such as requirements with associated tests. See also horizontal traceability, vertical traceability. |
| **trustworthiness** | The probability that there are no defects in the software that will cause the system to fail catastrophically. [Burnstein] |
| **usability** | The capability of the software to be understood, learned, used and attractive to the user when used under specified conditions. [ISO 9126] |
| **use case testing** | A black box test design technique in which test cases are designed to execute user scenarios. |
| **V-model** | A framework to describe the software development lifecycle activities from requirements specification to maintenance. The V-model illustrates how testing activities can be integrated into each phase of the software development lifecycle. |
| **validation** | Confirmation by examination and through provision of objective evidence that the requirements for a specific intended use or application have been fulfilled. [ISO 9000] |
| **verification** | Confirmation by examination and through provision of objective evidence that specified requirements have been fulfilled. [ISO 9000] |
| **vertical traceability** | The tracing of requirements through the layers of development documentation to components. |
| **walkthrough** | A step-by-step presentation by the author of a document in order to gather information and to establish a common understanding of its content. [Freedman and Weinberg, IEEE 1028] See also peer review. |
| **white-box test design technique** | Procedure to derive and/or select test cases based on an analysis of the internal structure of a component or system. |
| **white-box testing** | Testing based on an analysis of the internal structure of the component or system. |
| **Wide Band Delphi** | An expert based test estimation technique that aims at making an accurate estimation using the collective wisdom of the team members. |

# Appendix IV: ISTQB glossary terms and their classification

*This Appendix contains all terms, synonyms and definitions of the ISQTB software testing glossary. Also, we included the classification of all terms considering the 8 categories mentioned in Appendix I.*

| **ISTQB Glossary v3.5** | **ISTQB Terms Classified into 8 Categories or Conceptual Blocks for Testing.** Note that we include the synonyms of the terms in parentheses and not in bold. |
|---|---|
| **Cat#1 (C1): Test Project-, Strategy-, Organizational Test-related Terms** | |
| **Agile testing; analytical test strategy; bug hunting; capture/playback** (capture/replay; record/playback)**; Computer Aided Software Testing (CAST); consultative test strategy** (directed test strategy); **continuous testing; crowd testing; exhaustive testing** (complete testing)**; independence of testing; independent test lab (ITL); methodical test strategy; model-based test strategy; offline MBT; online MBT** (on-the-fly MBT)**; process-compliant test strategy; load management; PRISMA; reactive test strategy; regression-averse test strategy; requirements-based testing; session-based testing; standard-compliant test strategy; test approach; test infrastructure; test management; test phase; test policy** (organizational test policy); **test process group (TPG); test process improvement; test schedule; test automation strategy; test strategy** (organizational test strategy); **Systematic Test and Evaluation Process (STEP)**; **Test Maturity Model integration (TMMi)**; **test mission**; **TPI Next; exploratory testing** | |
| **Cat#2 (C2): Testing Work Process-, Activity-related Terms** | |
| **risk-based testing; acceptance testing; accessibility testing; alpha testing; API testing; audio testing; back-to-back-testing; beta testing; capacity testing; change-related testing; CLI testing; compliance testing** (conformance testing; regulation testing; standards testing); **component integration testing** (link testing); **component testing** (unit testing; module testing); **concurrency testing; confirmation testing** (re-testing)**; contractual acceptance testing; control flow analysis; Critical Testing Processes (CTP); data flow analysis; device-based testing; dynamic testing** (XiL test environment)**; end-to-end testing** (E2E testing)**; endurance testing; experience-based testing; expert usability review; field testing; formal review; functional testing; GUI testing; hardware in the loop (HiL); fault injection; fault seeding** (error seeding; bebugging)**; informal review; insourced testing; inspection; integration testing; interface testing; interoperability testing; load testing; maintenance testing; math testing; model in the loop (MiL); model-based testing (MBT); multiplayer testing; negative testing** (invalid testing; dirty testing); **neighborhood integration testing; non-functional testing; operational acceptance testing** (production acceptance testing)**; outsourced testing; pairwise integration testing; par sheet testing; peer review; performance testing; player perspective testing; post-release testing; proximity-based testing; malware scanning; static analysis; pair testing; reactive testing; regression testing; regulatory acceptance testing; review; scalability testing; scripted testing; security testing; software in the loop (SiL); software qualification test; spike testing; static code analysis; static testing; stress testing; suitability testing; system integration testing; system qualification test; system testing; technical review; test analysis; test completion; test control; test cycle; test data preparation; test design; test execution; test implementation; test level** (test stage); **test logging** (test recording); **test monitoring; test planning; test process; test session; test type; testing; usability test session; usability test task; usability testing; user acceptance testing (UAT); user-agent based testing; white-box testing** (clear-box testing; code-based testing; glass-box testing; logic-coverage testing; logic-driven testing; structural testing; structure-based testing); **test closure**; **test reporting**; **test run** | |
| **Cat#3 (C3): Test Goal-, Requirements-, Entity-related Terms** | |
| **Component** (Module; unit); **Driver** (test driver); **stub**; **system under test (SUT); test environment** (test bed; test rig); **test harness; test item; test object; test objective; test condition** (test situation; test requirement**)** | |
| **Cat#4 (C4): Test Work Product-related Terms (e.g. Artifacts, Reports, Results, Specifications)** | |
| **actual result** (actual outcome)**; automation code defect density; boundary value coverage; branch coverage; build verification test (BVT); classification tree; code coverage; condition coverage** (branch condition coverage)**; convergence metric; Coverage** (test coverage); **coverage item; decision coverage; defect density** (fault density)**; Defect Detection Percentage (DDP)** (Fault Detection Percentage (FDP)); **defect report** (bug report); **equivalence partition coverage**; **expected result** (predicted outcome; expected outcome); **false-negative result** (false-pass result); **false-positive result** (false-fail result); **high-level test** | |

| |
|---|
| **case** (logical test case; abstract test case); **Incident** (test incident; software test incident; deviation); **incident report** (software test incident report; test incident report; deviation report); **equivalent manual test effort (EMTE)**; **failed**; **level test plan; low-level test case** (concrete test case); **master test plan**; **MBT model**; **modified condition / decision coverage (MC/DC)** (modified multiple condition coverage; MC/DC coverage; condition determination coverage); **multiple condition coverage** (branch condition combination coverage; condition combination coverage); **N-switch coverage** (Chow's coverage metrics); **path coverage; load profile; model coverage; passed; review plan; smoke test** (sanity test; intake test; confidence test); **statement coverage; structural coverage; test; test basis; test case; test case specification; test charter** (charter)**; test data; test design specification; test estimation; test execution schedule; test improvement plan; test log** (test record; test run log); **test model; test plan; test progress report** (test status report); **test report; test result** (outcome; test outcome; result); **test script; test specification; test suite** (test set; test case suite); **test summary report; testware; usability test script; test Oracle** (oracle)**; tour** |
| **Cat#5 (C5): Testing Method-, Technique-related, Procedure-related, Rule-related Terms** |
| **ad hoc review; authentication; black-box test technique** (specification-based technique; black-box technique; specification-based test technique)**; boundary value analysis; cause-effect graphing** (cause-effect analysis); **checklist-based review; checklist-based testing; classification tree technique** (classification tree method)**; combinatorial testing; control flow testing; data-driven testing; decision condition testing; decision table testing; decision testing; defect-based test technique** (defect-based test design technique; defect-based technique); **equivalence partitioning** (partition testing); **error guessing**; **experience-based test technique** (experience-based test design technique; experience-based technique); **fuzz testing** (fuzzing); **Fault Tree Analysis (FTA)** (Software Fault Tree Analysis); **keyword-driven testing** (action word-driven testing); **linear scripting; modified condition / decision testing** (modified multiple condition testing; condition determination testing); **multiple condition testing** (branch condition combination testing; condition combination testing); **pairwise testing; pass/fail criteria; path testing; penetration testing; perspective-based Reading** (perspective-based reviewing); **process-driven scripting; random testing; role-based reviewing; scenario-based review; session-based test management (SBTM); state transition testing** (finite state testing)**; statement testing; structured scripting; Test Point Analysis (TPA); test technique** (test design technique; test specification technique; test case design technique); **think aloud usability testing; use case testing** (user scenario testing; scenario testing)**; user story testing; white-box test technique** (structure-based technique; structure-based test technique; structural test technique; white-box technique); **test procedure**; **test selection criteria**; **Walkthrough** (structured walkthrough) |
| **Cat#6 (C6): Testing Agent-, Role-,Tool-related Terms** |
| **capture/playback tool** (capture/replay tool; record/playback tool); **ethical hacker**; **hyperlink test tool; incident management tool; fault seeding tool** (error seeding tool)**; load generator; Moderator** (facilitator)**; monitoring tool; performance testing tool; Reviewer** (Checker; inspector); **Scribe** (recorder); **simulator; Software Usability Measurement Inventory (SUMI); static analyzer** (static analysis tool; analyzer)**; test automation engineer; test automation framework; test automation manager; test data preparation tool** (test generator); **test director; test execution tool; test leader** (lead tester); **test management tool; test manager; test process improver; test tool; tester; unit test framework; usability test participant; vulnerability scanner; test architect; test automation; test execution automation** |
| **Cat#7 (C7): Other Terms somewhat related to Test (e.g., anomaly, defect, etc.)** |
| **abnormal end** (abnormal termination); **abuse case**; **acceptance criteria**; **account harvesting; anomaly; anti-malware; attack vector; attacker; audit; boundary value; cause-effect graph; code injection; codependent behavior; computer forensics; content-based model** (content reference model)**; continuous integration; cross-site scripting (XSS); defect** (bug; fault)**; defect management; defect management committee** (defect triage committee)**; defect taxonomy** (bug taxonomy)**; denial of service (DoS); entry criteria** (definition of ready); **exit criteria** (test completion criteria; completion criteria; definition of done); **failure**; **mean time between failures (MTBF)**; **equivalence partition** (equivalence class); **escaped defect**; **failure mode**; **Failure Mode and Effect Analysis (FMEA)** (Software Failure Mode and Effect Analysis); **failure rate**; **generic test automation architecture**; **product risk**; **project risk**; **level of intrusion**; **method table; phase containment**; **postcondition; precondition**; **risk; risk impact** (impact); **risk level** (risk exposure); **risk likelihood** (likelihood); **risk type** (risk category); **remote test lab**; **test adaptation layer**; **test automation architecture; test automation solution**; **test case explosion**; **test definition layer**; **test execution layer; test generation layer; test hook; test process improvement manifesto; test pyramid; testing quadrants; usability lab** |
| **Cat#8 (C8): Terms beyond to the Test Domain related to Quality and Software Engineering** |

**acceptance test-driven development (ATDD); accessibility; accountability; acting (IDEAL); adaptability; Agile Manifesto; Agile software development; analyzability; anti-pattern; Appropriateness recognizability** (understandability); **assessment report; Assessor; atomic condition; authenticity; authorization; automotive safety integrity level (ASIL); automotive SPICE (ASPICE); availability; balanced scorecard; behavior-driven development (BDD); botnet; branch; call graph; Capability Maturity Model Integration (CMMI); capacity; cause-effect diagram** (fishbone diagram; Ishikawa diagram); **certification; change management; closed-loop-system; co-existence; coding standard; command-line interface (CLI); commercial off-the-shelf (COTS)** (off-the-shelf software)**; compatibility; compiler; complexity; compliance; concurrency; confidence interval; confidentiality; configuration item; configuration management; connectivity; context of use; continuous representation; control chart** (Shewhart chart)**; control flow; corporate dashboard; cost of quality; critical success factor; cross-browser compatibility; custom tool; cyclomatic complexity** (cyclomatic number)**; dashboard; data obfuscation; data privacy; debugging; debugging tool** (debugger); **decision; definition-use pair; demilitarized zone (DMZ); Deming cycle; diagnosing (IDEAL); dynamic analysis; effectiveness; efficiency; mean time to repair (MTTR)**; **emulator**; **encryption**; **environment model**; **epic**; **Error** (Mistake); **establishing (IDEAL)**; **European Foundation for Quality Management excellence model (EFQM)**; **Extreme Programming (XP)**; **failover**; **feature-driven development**; **firewall**; **formative evaluation**; **functional appropriateness** (suitability); **functional completeness**; **functional correctness** (accuracy); **functional safety** (safety); **functional suitability** (functionality); **Goal Question Metric (GQM)**; **graphical user interface (GUI)**; **hacker**; **hashing**; **heuristic**; **heuristic evaluation**; **human-centered design**; **hyperlink**; **IDEAL**; **impact analysis**; **emotional intelligence**; **error tolerance**; **fault tolerance**; **finding**; **incremental development model**; **indicator**; **information assurance**; **initiating (IDEAL)**; **insider threat**; **installability; integrity; interoperability**; **intrusion detection system (IDS); iterative development model**; **lead assessor; learnability; learning (IDEAL)**; **maintainability; maintenance**; **malware**; **man-in-the-middle attack**; **management review**; **manufacturing-based quality**; **maturity; maturity level; maturity model**; **measure; measurement; memory leak**; **metric; mind map** (Mind-Map); **Modifiability** (changeability); **modularity**; **Myers-Briggs Type Indicator (MBTI)**; **network zone**; **non-repudiation**; **open-source tool** (open source tool); **open–loop-system; operability**; **operational profile; operational profiling**; **Pareto analysis**; **password cracking; Path** (control flow path); **performance efficiency; performance indicator** (key performance indicator); **pharming**; **phishing; planning poker**; **pointer; portability**; **priority**; **process assessment; process model; product-based quality; load generation; risk analysis; risk assessment; risk identification; risk management; risk mitigation; probe effect; quality; quality assurance (QA); quality characteristic** (quality attribute; software quality characteristic; software product characteristic)**; quality control (QC); quality function deployment (QFD); quality management; quality risk; RACI matrix; ramp-down; ramp-up; Rational Unified Process (RUP); Reconnaissance** (footprinting)**; recoverability; reliability; reliability growth model; replaceability; requirement; resource utilization** (storage)**; retrospective meeting** (post-project meeting; project retrospective)**; reusability; robustness; root cause; root cause analysis** (causal analysis)**; S.M.A.R.T. goal methodology (SMART); salting; scalability; script kiddie; scrum; security** (information security); **security attack; security audit; security policy; security procedure; security risk; security vulnerability; sequential development model; service virtualization; severity; short-circuiting; social engineering; software development lifecycle (SDLC); software lifecycle; software process improvement (SPI); specification by example (SBE); SQL injection; staged representation; standard; Statement** (source statement)**; summative evaluation; system hardening; system of systems** (multi-system)**; system throughput; System Usability Scale (SUS); test-driven development (TDD); test-first approach; think time; time behavior; Total Quality Management (TQM); traceability; traceability matrix; transactional analysis; transcendent-based quality; usability; usability evaluation; usability requirement; user error protection; user experience; user interface; user interface aesthetics** (attractiveness)**; user interface guideline; user story; user survey; user-based quality; V-model; validation; value-based quality; verification; virtual user; Web Content Accessibility Guidelines (WCAG); Website Analysis and Measurement Inventory (WAMMI); Wideband Delphi; wild pointer; testability**

| **Term name**<br>synonyms | **Definition** |
|---|---|
| **abnormal end**<br>abnormal termination | The unintended termination of the execution of a component or system prior to completion. |
| **abuse case** | A use case in which some actors with malicious intent are causing harm to the system or to other actors. |
| **acceptance criteria** | The criteria that a component or system must satisfy in order to be accepted by a user, customer, or other authorized entity. |
| **acceptance test-driven development (ATDD)** | A collaborative approach to development in which the team and customers are using the customers own domain language to understand their requirements, which forms the basis for testing a component or system. |
| **acceptance testing** | A test level that focuses on determining whether to accept the system. |
| **accessibility** | The degree to which a component or system can be used by people with the widest range of characteristics and capabilities to achieve a specified goal in a specified context of use. |
| **accessibility testing** | Testing to determine the ease by which users with disabilities can use a component or system. |
| **account harvesting** | The process of obtaining user account information based on trial and error with the intention of using that information in a security attack. |
| **accountability** | The degree to which the actions of an entity can be traced uniquely to that entity. |
| **acting (IDEAL)** | The phase within the IDEAL model where the improvements are developed, put into practice, and deployed across the organization. The acting phase consists of the activities: create solution, pilot/test solution, refine solution and implement solution. |
| **actual result**<br>actual outcome | The behavior produced/observed when a component or system is tested. |
| **ad hoc review** | A review technique performed informally without a structured process. |
| **adaptability** | The degree to which a component or system can be adapted for different or evolving hardware, software or other operational or usage environments. |
| **Agile Manifesto** | A statement on the values that underpin Agile software development. The values are: individuals and interactions over processes and tools, working software over comprehensive documentation, customer collaboration over contract negotiation, responding to change over following a plan. |
| **Agile software development** | A group of software development methodologies based on iterative incremental development, where requirements and solutions evolve through collaboration between self-organizing cross-functional teams. |
| **Agile testing** | Testing practice for a project using Agile software development methodologies, incorporating techniques and methods, such as extreme programming (XP), treating development as the customer of testing and emphasizing the test-first design paradigm. |
| **alpha testing** | A type of acceptance testing performed in the developer's test environment by roles outside the development organization. |
| **analytical test strategy** | A test strategy whereby the test team analyzes the test basis to identify the test conditions to cover. |
| **analyzability** | The degree to which an assessment can be made for a component or system of either the impact of one or more intended changes, the diagnosis of deficiencies or causes of failures, or the identification of parts to be modified. |
| **anomaly** | Any condition that deviates from expectation based on requirements specifications, design documents, user documents, standards, etc., or from someone's perception or experience. Anomalies may be found during, but not limited to, reviewing, testing, analysis, compilation, or use of software products or applicable documentation. |
| **anti-malware** | Software that is used to detect and inhibit malware. |
| **anti-pattern** | Repeated action, process, structure or reusable solution that initially appears to be beneficial and is commonly used but is ineffective and/or counterproductive in practice. |
| **API testing** | Testing performed by submitting commands to the software under test using programming interfaces of the application directly. |
| **Appropriateness recognizability**<br>understandability | The degree to which users can recognize whether a component or system is appropriate for their needs. |
| **assessment report** | A document summarizing the assessment results, e.g., conclusions, recommendations and findings. |
| **assessor** | A person who conducts an assessment. Any member of an assessment team. |
| **atomic condition** | A condition that does not contain logical operators. |
| **attack vector** | A path or means by which an attacker can gain access to a system for malicious purposes. |
| **attacker** | A person or process that attempts to access data, functions or other restricted areas of the system without authorization, potentially with malicious intent. |
| **audio testing** | Testing to determine if the game music and sound effects will engage the user in the game and enhance the game play. |
| **audit** | An independent examination of a work product or process performed by a third party to assess whether it complies with specifications, standards, contractual agreements, or other criteria. |
| **authentication** | A procedure determining whether a person or a process is, in fact, who or what it is declared to be. |
| **authenticity** | The degree to which the identity of a subject or resource can be proved to be the one claimed. |
| **authorization** | Permission given to a user or process to access resources. |
| **automation code defect density** | Defect density of a component of the test automation code. |
| **automotive safety integrity level (ASIL)** | One of four levels that specify the item's or element's necessary requirements of ISO 26262 and safety measures to avoid an unreasonable residual risk. |
| **automotive SPICE (ASPICE)** | A process reference model and an associated process assessment model in the automotive industry that conforms with the requirements of ISO/IEC 33002:2015. |
| **availability** | The degree to which a component or system is operational and accessible when required for use. |
| **back-to-back-testing** | Testing to compare two or more variants of a test item or a simulation model of the same test item by executing the same test cases on all variants and comparing the results. |
| **balanced scorecard** | A strategic tool for measuring whether the operational activities of a company are aligned with its objectives in terms of business vision and strategy. |
| **behavior-driven development (BDD)** | A collaborative approach to development in which the team is focusing on delivering expected behavior of a component or system for the customer, which forms the basis for testing. |
| **beta testing** | A type of acceptance testing performed at an external site to the developer's test environment by roles outside the development organization. |
| **black-box test technique** specification-based technique; black-box technique; specification-based test technique | A test technique based on an analysis of the specification of a component or system. |
| **botnet** | A network of compromised computers, called bots or robots, which is controlled by a third party and used to transmit malware or spam, or to launch attacks. |
| **boundary value** | A minimum or maximum value of an ordered equivalence partition. |
| **boundary value analysis** | A black-box test technique in which test cases are designed based on boundary values. |
| **boundary value coverage** | The coverage of boundary values. |
| **branch** | A transfer of control from a decision point. |
| **branch coverage** | The coverage of branches. |
| **bug hunting** | An approach to testing in which gamification and awards for defects found are used as a motivator. |
| **build verification test (BVT)** | A set of automated tests which validates the integrity of each new build and verifies its key/core functionality, stability and testability. |
| **call graph** | An abstract representation of calling relationships between subroutines in a program. |
| **Capability Maturity Model Integration (CMMI)** | A framework that describes the key elements of an effective product development and maintenance process. The Capability Maturity Model Integration covers best-practices for planning, engineering and managing product development and maintenance. |
| **capacity** | The degree to which the maximum limits of a component or system parameter meet requirements. |
| **capacity testing** | Testing to evaluate the capacity of a system. |
| **capture/playback** capture/replay; | A test automation approach in which inputs to the test object are recorded during manual testing to generate automated test scripts that can be executed later. |
| record/playback | |
| **capture/playback tool** capture/replay tool; record/playback tool | A type of test execution tool where inputs are recorded during manual testing in order to generate automated test scripts that can be executed later (i.e. replayed). These tools are often used to support automated regression testing. |
| **cause-effect diagram** fishbone diagram; Ishikawa diagram | A graphical representation used to organize and display the interrelationships of various possible root causes of a problem. Possible causes of a real or potential defect or failure are organized in categories and subcategories in a horizontal tree-structure, with the (potential) defect or failure as the root node. |
| **cause-effect graph** | A graphical representation of logical relationships between inputs (causes) and their associated outputs (effects). |
| **cause-effect graphing** cause-effect analysis | A black-box test technique in which test cases are designed from cause-effect graphs. |
| **certification** | The process of confirming that a component, system or person complies with its specified requirements. |
| **change management** | (1) A structured approach to transitioning individuals and organizations from a current state to a desired future state. (2) Controlled way to effect a change, or a proposed change, to a product or service. |
| **change-related testing** | A type of testing initiated by modification to a component or system. |
| **checklist-based review** | A review technique guided by a list of questions or required attributes. |
| **checklist-based testing** | An experience-based test technique whereby the experienced tester uses a high-level list of items to be noted, checked, or remembered, or a set of rules or criteria against which a product has to be verified. |
| **classification tree** | A tree diagram representing test data domains of a test object. |
| **classification tree technique** classification tree method | A black-box test technique in which test cases are designed using a classification tree. |
| **CLI testing** | Testing performed by submitting commands to the software under test using a dedicated command-line interface. |
| **closed-loop-system** | A system in which the controlling action or input is dependent on the output or changes in output. |
| **co-existence** | The degree to which a component or system can perform its required functions while sharing an environment and resources with other components or systems without a negative impact on any of them. |
| **code coverage** | The coverage of code. |
| **code injection** | A type of security attack performed by inserting malicious code at an interface into an application to exploit poor handling of untrusted data. |
| **codependent behavior** | Excessive emotional or psychological dependence on another person, specifically in trying to change that person's current (undesirable) behavior while supporting them in continuing that behavior. For example, in software testing, complaining about late delivery to test and yet enjoying the necessary "heroism", working additional hours to make up time when delivery is running late, therefore reinforcing the lateness. |
| **coding standard** | A standard that describes the characteristics of a design or a design description of data or program components. |
| **combinatorial testing** | A black-box test technique in which test cases are designed to exercise specific combinations of values of several parameters |
| **command-line interface (CLI)** | A type of interface in which the information is passed in form of command lines. |
| **commercial off-the-shelf (COTS)** off-the-shelf software | A type of product developed in an identical format for a large number of customers in the general market. |
| **compatibility** | The degree to which a component or system can exchange information with other components or systems, and/or perform its required functions while sharing the same hardware or software environment. |
| **compiler** | A software tool that translates programs expressed in a high-order language into their machine language equivalents. |
| **complexity** | The degree to which a component or system has a design and/or internal structure that is difficult to understand, maintain and verify. |
| **compliance** | Adherence of a work product to standards, conventions or regulations in laws and similar prescriptions. |
| **compliance testing**<br>conformance testing;<br>regulation testing;<br>standards testing | Testing to determine the compliance of the component or system. |
| **Component**<br>Module;<br>unit | A part of a system that can be tested in isolation. |
| **component integration testing**<br>link testing | Testing in which the test items are interfaces and interactions between integrated components. |
| **component testing**<br>unit testing;<br>module testing | A test level that focuses on individual hardware or software components. |
| **Computer Aided Software Testing (CAST)** | The computing-based processes, techniques, and tools to support testing. |
| **computer forensics** | The practice of determining how a security attack has succeeded and assessing the damage caused. |
| **concurrency** | The simultaneous execution of multiple independent threads by a component or system. |
| **concurrency testing** | Testing to evaluate if a component or system involving concurrency behaves as specified. |
| **condition coverage**<br>branch condition coverage | The coverage of condition outcomes. |
| **confidence interval** | In managing project risks, the period of time within which a contingency action must be implemented in order to be effective in reducing the impact of the risk. |
| **confidentiality** | The degree to which a component or system ensures that data are accessible only to those authorized to have access. |
| **configuration item** | An aggregation of work products that is designated for configuration management and treated as a single entity in the configuration management process. |
| **configuration management** | A discipline applying technical and administrative direction and surveillance to identify and document the functional and physical characteristics of a configuration item, control changes to those characteristics, record and report change processing and implementation status, and verify that it complies with specified requirements. |
| **confirmation testing**<br>re-testing | A type of change-related testing performed after fixing a defect to confirm that a failure caused by that defect does not reoccur. |
| **connectivity** | The degree to which a component or system can connect to other components or systems. |
| **consultative test strategy**<br>directed test strategy | A test strategy whereby the test team relies on the input of one or more key stakeholders to determine the details of the strategy. |
| **content-based model**<br>content reference model | A process model providing a detailed description of good engineering practices, e.g., test practices. |
| **context of use** | Users, tasks, equipment (hardware, software and materials), and the physical and social environments in which a software product is used. |
| **continuous integration** | An automated software development procedure that merges, integrates and tests all changes as soon as they are committed. |
| **continuous representation** | A capability maturity model structure wherein capability levels provide a recommended order for approaching process improvement within specified process areas. |
| **continuous testing** | An approach that involves a process of testing early, testing often, test everywhere, and automate to obtain feedback on the business risks associated with a software release candidate as rapidly as possible. |
| **contractual acceptance testing** | A type of acceptance testing performed to verify whether a system satisfies its contractual requirements. |
| **control chart**<br>Shewhart chart | A statistical process control tool used to monitor a process and determine whether it is statistically controlled. It graphically depicts the average value and the upper and lower control limits (the highest and lowest values) of a process. |
| **control flow** | The sequence in which operations are performed by a business process, component or system. |
| **control flow analysis** | A type of static analysis based on a representation of unique paths for executing a component or system. |
| **control flow testing** | A white-box test technique in which test cases are designed based on control flows. |
| **convergence metric** | A metric that shows progress toward a defined criterion, e.g., convergence of the total number of tests executed to the total number of tests planned for execution. |
| **corporate dashboard** | A dashboard-style representation of the status of corporate performance data. |
| **cost of quality** | The total costs incurred on quality activities and issues and often split into prevention costs, appraisal costs, internal failure costs and external failure costs. |
| **Coverage**<br>test coverage | The degree to which specified coverage items have been determined or have been exercised by a test suite expressed as a percentage. |
| **coverage item** | An attribute or combination of attributes derived from one or more test conditions by using a test technique. |
| **critical success factor** | An element necessary for an organization or project to achieve its mission. Critical success factors are the critical factors or activities required for ensuring the success. |
| **Critical Testing Processes (CTP)** | A content-based model for test process improvement built around twelve critical processes. These include highly visible processes, by which peers and management judge competence and mission-critical processes in which performance affects the company's profits and reputation. |
| **cross-browser compatibility** | The degree to which a website or web application can function across different browsers and degrade gracefully when browser features are absent or lacking. |
| **cross-site scripting (XSS)** | A vulnerability that allows attackers to inject malicious code into an otherwise benign website. |
| **crowd testing** | An approach to testing in which testing is distributed to a large group of testers. |
| **custom tool** | A software tool developed specifically for a set of users or customers. |
| **cyclomatic complexity**<br>cyclomatic number | The maximum number of linear, independent paths through a program. |
| **dashboard** | A representation of dynamic measurements of operational performance for some organization or activity, using metrics represented via metaphors such as visual dials, counters, and other devices resembling those on the dashboard of an automobile, so that the effects of events or activities can be easily understood and related to operational goals. |
| **data flow analysis** | A type of static analysis based on the lifecycle of variables. |
| **data obfuscation** | Data transformation that makes it difficult for a human to recognize the original data. |
| **data privacy** | The protection of personally identifiable information or otherwise sensitive information from undesired disclosure. |
| **data-driven testing** | A scripting technique that uses data files to contain the test data and expected results needed to execute the test scripts. |
| **debugging** | The process of finding, analyzing and removing the causes of failures in a component or system. |
| **debugging tool**<br>debugger | A tool used by programmers to reproduce failures, investigate the state of programs and find the corresponding defect. Debuggers enable programmers to execute programs step by step, to halt a program at any program statement and to set and examine program variables. |
| **decision** | A type of statement in which a choice between two or more possible outcomes controls which set of actions will result. |
| **decision condition testing** | A white-box test technique in which test cases are designed to execute condition outcomes and decision outcomes. |
| **decision coverage** | The coverage of decision outcomes. |
| **decision table testing** | A black-box test technique in which test cases are designed to exercise the combinations of conditions and the resulting actions shown in a decision table. |
| **decision testing** | A white-box test technique in which test cases are designed to execute decision outcomes. |
| **Defect**<br>bug, fault | An imperfection or deficiency in a work product where it does not meet its requirements or specifications. |
| **defect density**<br>fault density | The number of defects per unit size of a work product. |
| **Defect Detection Percentage (DDP)**<br>Fault Detection Percentage (FDP) | The number of defects found by a test level, divided by the number found by that test level and any other means afterwards. |
| **defect management** | The process of recognizing, recording, classifying, investigating, resolving and disposing of defects. |
| **defect management committee**<br>defect triage committee | A cross-functional team of stakeholders who manage reported defects from initial detection to ultimate resolution (defect removal, defect deferral, or report cancellation). In some cases, the same team as the configuration control board. |
| **defect report**<br>bug report | Documentation of the occurrence, nature, and status of a defect. |
| **defect taxonomy**<br>bug taxonomy | A list of categories designed to identify and classify defects. |
| **defect-based test technique**<br>defect-based test design technique; defect-based technique | A test technique in which test cases are developed from what is known about a specific defect type. |
| **definition-use pair** | The association of a definition of a variable with the subsequent use of that variable. |
| **demilitarized zone (DMZ)** | A physical or logical subnetwork that contains and exposes an organization's external-facing services to an untrusted network, commonly the Internet. |
| **Deming cycle** | An iterative four-step problem-solving process (plan-do-check-act) typically used in process improvement. |
| **denial of service (DoS)** | A security attack that is intended to overload the system with requests such that legitimate requests cannot be serviced. |
| **device-based testing** | A type of testing in which test suites are executed on physical or virtual devices. |
| **diagnosing (IDEAL)** | The phase within the IDEAL model where it is determined where one is, relative to where one wants to be. The diagnosing phase consists of the activities to characterize current and desired states and develop recommendations. |
| **Driver**<br>test driver | A temporary component or tool that replaces another component and controls or calls a test item in isolation. |
| **dynamic analysis** | The process of evaluating a component or system based on its behavior during execution. |
| **dynamic testing** | Testing that involves the execution of the test item. |
| **effectiveness** | Extent to which correct and complete goals are achieved. |
| **efficiency** | The degree to which resources are expended in relation to results achieved. |
| **emotional intelligence** | The ability, capacity, and skill to identify, assess, and manage the emotions of one's self, of others, and of groups. |
| **emulator** | A device, computer program, or system that accepts the same inputs and produces the same outputs as a given system. |
| **encryption** | The process of encoding information so that only authorized parties can retrieve the original information, usually by means of a specific decryption key or process. |
| **end-to-end testing**<br>E2E testing | A type of testing in which business processes are tested from start to finish under production-like circumstances. |
| **endurance testing** | Testing to determine the stability of a system under a significant load over a significant period of time within the system's operational context. |
| **entry criteria**<br>definition of ready | The set of conditions for officially starting a defined task. |
| **environment model** | An abstraction of the real environment of a component or system including other components, processes, and environment conditions, in a real-time simulation. |
| **epic** | A large user story that cannot be delivered as defined within a single iteration or is large enough that it can be split into smaller user stories. |
| **equivalence partition**<br>equivalence class | A subset of the value domain of a variable within a component or system in which all values are expected to be treated the same based on the specification. |
| **equivalence partition coverage** | The coverage of equivalence partitions. |
| **equivalence partitioning**<br>partition testing | A black-box test technique in which test cases are designed to exercise equivalence partitions by using one representative member of each partition. |
| **equivalent manual test effort (EMTE)** | Effort required for running tests manually. |
| **Error**<br>mistake | A human action that produces an incorrect result. |
| **error guessing** | A test technique in which tests are derived on the basis of the tester's knowledge of past failures, or general knowledge of failure modes. |
| **error tolerance** | The degree to which a component or system can continue normal operation despite the presence of erroneous inputs. |
| **escaped defect** | A defect that was not detected by a testing activity that is supposed to find that defect. |
| **establishing (IDEAL)** | The phase within the IDEAL model where the specifics of how an organization will reach its destination are planned. The establishing phase consists of the activities set priorities, develop approach and plan actions. |
| **ethical hacker** | A security tester using hacker techniques. |
| **European Foundation for Quality Management excellence model (EFQM)** | A non-prescriptive framework for an organization's quality management system, defined and owned by the European Foundation for Quality Management, based on five 'Enabling' criteria (covering what an organization does), and four 'Results' criteria (covering what an organization achieves). |
| **exhaustive testing**<br>complete testing | A test approach in which the test suite comprises all combinations of input values and preconditions. |
| **exit criteria**<br>test completion criteria;<br>completion criteria;<br>definition of done | The set of conditions for officially completing a defined task. |
| **expected result**<br>predicted outcome;<br>expected outcome | The observable predicted behavior of a test item under specified conditions based on its test basis. |
| **experience-based test technique**<br>experience-based test design technique;<br>experience-based technique | A test technique only based on the tester's experience, knowledge and intuition. |
| **experience-based testing** | Testing based on the tester's experience, knowledge and intuition. |
| **expert usability review** | An informal usability review in which the reviewers are experts. Experts can be usability experts or subject matter experts, or both. |
| **exploratory testing** | An approach to testing whereby the testers dynamically design and execute tests based on their knowledge, exploration of the test item and the results of previous tests. |
| **Extreme Programming (XP)** | A software engineering methodology used within Agile software development whereby core practices are programming in pairs, doing extensive code review, unit testing of all code, and simplicity and clarity in code. |
| **failed** | The status of a test result in which the actual result does not match the expected result. |
| **failover** | The backup operational mode in which the functions of a system that becomes unavailable are assumed by a secondary system. |
| **failure** | An event in which a component or system does not perform a required function within specified limits. |
| **failure mode** | The physical or functional manifestation of a failure. |
| **Failure Mode and Effect Analysis (FMEA)** Software Failure Mode and Effect Analysis | A systematic approach to risk identification and analysis of identifying possible modes of failure and attempting to prevent their occurrence. |
| **failure rate** | The ratio of the number of failures of a given category to a given unit of measure. |
| **false-negative result** false-pass result | A test result which fails to identify the presence of a defect that is actually present in the test object. |
| **false-positive result** false-fail result | A test result in which a defect is reported although no such defect actually exists in the test object. |
| **fault injection** | The process of intentionally adding defects to a system for the purpose of finding out whether the system can detect, and possibly recover from, a defect. Fault injection is intended to mimic failures that might occur in the field. |
| **fault seeding** error seeding; bebugging | The process of intentionally adding known faults to those already in a component or system to monitor the rate of detection and removal, and to estimate the number of faults remaining. |
| **fault seeding tool** error seeding tool | A tool for seeding (i.e., intentionally inserting) faults in a component or system. |
| **fault tolerance** | The degree to which a component or system operates as intended despite the presence of hardware or software faults. |
| **Fault Tree Analysis (FTA)** Software Fault Tree Analysis | A technique used to analyze the causes of faults (defects). The technique visually models how logical relationships between failures, human errors, and external events can combine to cause specific faults to disclose. |
| **feature-driven development** | An iterative and incremental software development process driven from a client-valued functionality (feature) perspective. Feature-driven development is mostly used in Agile software development. |
| **field testing** | A type of testing conducted to evaluate the system behavior under productive connectivity conditions in the field. |
| **finding** | A result of an evaluation that identifies some important issue, problem, or opportunity. |
| **firewall** | A component or set of components that controls incoming and outgoing network traffic based on predetermined security rules. |
| **formal review** | A type of review that follows a defined process with a formally documented output. |
| **formative evaluation** | A type of evaluation designed and used to improve the quality of a component or system, especially when it is still being designed. |
| **functional appropriateness** suitability | The degree to which the functions facilitate the accomplishment of specified tasks and objectives. |
| **functional completeness** | The degree to which the set of functions covers all the specified tasks and user objectives. |
| **functional correctness** accuracy | The degree to which a component or system provides the correct results with the needed degree of precision. |
| **functional safety** safety | The absence of unreasonable risk due to hazards caused by malfunctioning behavior of Electric/Electronic(E/E) - Systems. |
| **functional suitability** functionality | The degree to which a component or system provides functions that meet stated and implied needs when used under specified conditions. |
| **functional testing** | Testing performed to evaluate if a component or system satisfies functional requirements. |
| **fuzz testing**<br>fuzzing | A software testing technique used to discover security vulnerabilities by inputting massive amounts of random data, called fuzz, to the component or system. |
| **generic test automation architecture** | Representation of the layers, components, and interfaces of a test automation architecture, allowing for a structured and modular approach to implement test automation. |
| **Goal Question Metric (GQM)** | An approach to software measurement using a three-level model conceptual level (goal), operational level (question) and quantitative level (metric). |
| **graphical user interface (GUI)** | A type of interface that allows users to interact with a component or system through graphical icons and visual indicators. |
| **GUI testing** | Testing performed by interacting with the software under test via the graphical user interface. |
| **hacker** | A person or organization who is actively involved in security attacks, usually with malicious intent. |
| **hardware in the loop (HiL)** | Dynamic testing performed using real hardware with integrated software in a simulated environment. |
| **hashing** | Transformation of a variable length string of characters into a usually shorter fixed-length value or key. Hashed values, or hashes, are commonly used in table or database lookups. Cryptographic hash functions are used to secure data. |
| **heuristic** | A generally recognized rule of thumb that helps to achieve a goal. |
| **heuristic evaluation** | An evaluation of a work product that uses a heuristic. |
| **high-level test case**<br>logical test case;<br>abstract test case | A test case with abstract preconditions, input data, expected results, postconditions, and actions (where applicable). |
| **human-centered design** | An approach to design that aims to make software products more usable by focusing on the use of the software products and applying human factors, ergonomics, and usability knowledge and techniques. |
| **hyperlink** | A pointer within a web page that leads to other web pages. |
| **hyperlink test tool** | A tool used to check that no broken hyperlinks are present on a web site. |
| **IDEAL** | An organizational improvement model that serves as a roadmap for initiating, planning, and implementing improvement actions. The IDEAL model is named for the five phases it describes: initiating, diagnosing, establishing, acting, and learning. |
| **impact analysis** | The identification of all work products affected by a change, including an estimate of the resources needed to accomplish the change. |
| **Incident**<br>test incident;<br>software test incident;<br>deviation | An event occurring that requires investigation. |
| **incident management tool** | A tool that facilitates the recording and status tracking of incidents. |
| **incident report**<br>software test incident report;<br>test incident report;<br>deviation report | Documentation of the occurrence, nature, and status of an incident. |
| **incremental development model** | A type of software development lifecycle model in which the component or system is developed through a series of increments. |
| **independence of testing** | Separation of responsibilities, which encourages the accomplishment of objective testing. |
| **independent test lab (ITL)** | An organization responsible to test and certify that the software, hardware, firmware, platform, and operating system follow all the jurisdictional rules for each location where the product will be used. |
| **indicator** | A measure that provides an estimate or evaluation of specified attributes derived from a model with respect to defined information needs. |
| **informal review** | A type of review that does not follow a defined process and has no formally documented output. |
| **information assurance** | Measures that protect and defend information and information systems by ensuring their availability, integrity, authentication, confidentiality, and nonrepudiation. These measures include providing for restoration of information systems by incorporating protection, detection, and reaction capabilities. |
| **initiating (IDEAL)** | The phase within the IDEAL model where the groundwork is laid for a successful improvement effort. The initiating phase consists of the activities: set context, build sponsorship and charter infrastructure. |
| **insider threat** | A security threat originating from within the organization, often by an authorized system user. |
| **insourced testing** | Testing performed by people who are co-located with the project team but are not fellow employees. |
| **inspection** | A type of formal review to identify issues in a work product, which provides measurement to improve the review process and the software development process. |
| **installability** | The degree to which a component or system can be successfully installed and/or uninstalled in a specified environment. |
| **integration testing** | A test level that focuses on interactions between components or systems. |
| **integrity** | The degree to which a component or system allows only authorized access and modification to a component, a system or data. |
| **interface testing** | A type of integration testing performed to determine whether components or systems pass data and control correctly to one another. |
| **interoperability** | The degree to which two or more components or systems can exchange information and use the information that has been exchanged. |
| **interoperability testing** | Testing to determine the interoperability of a software product. |
| **intrusion detection system (IDS)** | A system which monitors activities on the 7 layers of the OSI model from network to application level, to detect violations of the security policy. |
| **iterative development model** | A type of software development lifecycle model in which the component or system is developed through a series of repeated cycles. |
| **keyword-driven testing** action word-driven testing | A scripting technique in which test scripts contain high-level keywords and supporting files that contain low-level scripts that implement those keywords. |
| **lead assessor** | The person who leads an assessment. In some cases, for instance CMMI and TMMi when formal assessments are conducted, the lead assessor must be accredited and formally trained. |
| **learnability** | The degree to which a component or system can be used by specified users to achieve specified goals of learning with satisfaction and freedom from risk in a specified context of use. |
| **learning (IDEAL)** | The phase within the IDEAL model where one learns from experiences and improves one's ability to adopt new processes and technologies in the future. The learning phase consists of the activities: analyze and validate, and propose future actions. |
| **level of intrusion** | The level to which a test object is modified by adjusting it for testability. |
| **level test plan** | A test plan that typically addresses one test level. |
| **linear scripting** | A simple scripting technique without any control structure in the test scripts. |
| **load generation** | The process of simulating a defined set of activities at a specified load to be submitted to a component or system. |
| **load generator** | A tool that generates a load for a system under test. |
| **load management** | The control and execution of load generation, and performance monitoring and reporting of the component or system. |
| **load profile** | Documentation defining a designated number of virtual users who process a defined set of transactions in a specified time period that a component or system being tested may experience in production. |
| **load testing** | A type of performance testing conducted to evaluate the behavior of a component or system under varying loads, usually between anticipated conditions of low, typical, and peak usage. |
| **low-level test case** concrete test case | A test case with concrete values for preconditions, input data, expected results and postconditions and detailed description of actions (where applicable). |
| **maintainability** | The degree to which a component or system can be modified by the intended maintainers. |
| **maintenance** | The process of modifying a component or system after delivery to correct defects, improve quality characteristics, or adapt to a changed environment. |
| **maintenance testing** | Testing the changes to an operational system or the impact of a changed environment to an operational system. |
| **malware** | Software that is intended to harm a system or its components. |
| **malware scanning** | Static analysis aiming to detect and remove malicious code received at an interface. |
| **man-in-the-middle attack** | The interception, mimicking and/or altering and subsequent relaying of communications (e.g., credit card transactions) by a third party such that a user remains unaware of that third party's presence. |
| **management review** | A systematic evaluation of software acquisition, supply, development, operation, or maintenance process, performed by or on behalf of management that monitors progress, determines the status of plans and schedules, confirms requirements and their system allocation, or evaluates the effectiveness of management approaches to achieve fitness for purpose. |
| **manufacturing-based quality** | A view of quality, whereby quality is measured by the degree to which a product or service conforms to its intended design and requirements. Quality arises from the process(es) used. |
| **master test plan** | A test plan that is used to coordinate multiple test levels or test types. |
| **math testing** | Testing to determine the correctness of the pay table implementation, the random number generator results, and the return to player computations. |
| **maturity** | (1) The capability of an organization with respect to the effectiveness and efficiency of its processes and work practices. (2) The degree to which a component or system meets needs for reliability under normal operation. |
| **maturity level** | Degree of process improvement across a predefined set of process areas in which all goals in the set are attained. |
| **maturity model** | A structured collection of elements that describe certain aspects of maturity in an organization, and aid in the definition and understanding of an organization's processes. |
| **MBT model** | Any model used in model-based testing. |
| **mean time between failures (MTBF)** | The average time between failures of a component or system. |
| **mean time to repair (MTTR)** | The average time a component or system will take to recover from a failure. |
| **measure** | The number or category assigned to an attribute of an entity by making a measurement. |
| **measurement** | The process of assigning a number or category to an entity to describe an attribute of that entity. |
| **memory leak** | A memory access failure due to a defect in a program's dynamic store allocation logic that causes it to fail to release memory after it has finished using it. |
| **method table** | A table containing different test approaches, testing techniques and test types that are required depending on the Automotive Safety Integrity Level (ASIL) and on the context of the test object. |
| **methodical test strategy** | A test strategy whereby the test team uses a pre-determined set of test conditions such as a quality standard, a checklist, or a collection of generalized, logical test conditions which may relate to a particular domain, application or type of testing. |
| **metric** | A measurement scale and the method used for measurement. |
| **mind map**<br>Mind-Map | A diagram arranged around a general theme that represents ideas, tasks, words or other items. |
| **model coverage** | The coverage of model elements. |
| **model in the loop (MiL)** | Dynamic testing performed using a simulation model of the system in a simulated environment. |
| **model-based test strategy** | A test strategy whereby the test team derives testware from models. |
| **model-based testing (MBT)** | Testing based on or involving models. |
| **Moderator**<br>facilitator | (1) The person responsible for running review meetings. (2) The person who conducts a usability test session. |
| **Modifiability**<br>changeability | The degree to which a component or system can be changed without introducing defects or degrading existing product quality. |
| **modified condition / decision coverage (MC/DC)** modified multiple condition coverage; MC/DC coverage; condition determination coverage | The coverage of all outcomes of the atomic conditions that independently affect the overall decision outcome. |
| **modified condition / decision testing** modified multiple condition testing; condition determination testing | A white-box test technique in which test cases are designed to exercise outcomes of atomic conditions that independently affect a decision outcome. |
| **modularity** | The degree to which a system is composed of discrete components such that a change to one component has minimal impact on other components. |
| **monitoring tool** | A software tool or hardware device that runs concurrently with the component or system under test and supervises, records and/or analyzes the behavior of the component or system. |
| **multiplayer testing** | Testing to determine if many players can simultaneously interact with the casino game world, with computer-controlled opponents, game servers, and with each other, as expected according to the game design. |
| **multiple condition coverage** branch condition combination coverage; condition combination coverage | The coverage of all possible combinations of all single condition outcomes within one statement. |
| **multiple condition testing** branch condition combination testing; condition combination testing | A white-box test technique in which test cases are designed to exercise outcome combinations of atomic conditions. |
| **Myers-Briggs Type Indicator (MBTI)** | An indicator of psychological preference representing the different personalities and communication styles of people. |
| **N-switch coverage** Chow's coverage metrics | The coverage of sequences of N+1 transitions. |
| **negative testing** invalid testing; dirty testing | Testing a component or system in a way for which it was not intended to be used. |
| **neighborhood integration testing** | A type of integration testing in which all of the nodes that connect to a given node are the basis for the integration testing. |
| **network zone** | A sub-network with a defined level of trust. For example, the Internet or a public zone would be considered to be untrusted. |
| **non-functional testing** | Testing performed to evaluate that a component or system complies with non-functional requirements. |
| **non-repudiation** | The degree to which actions or events can be proven to have taken place, so that the actions or events cannot be repudiated later. |
| **offline MBT** | Model-based test approach whereby test cases are generated into a repository for future execution. |
| **online MBT** on-the-fly MBT | Model-based test approach whereby test cases are generated and executed simultaneously. |
| **open-source tool** open source tool | A software tool that is available to all potential users in source code form, usually via the internet. Its users are permitted, usually under license, to study, change, improve and, at times, to distribute the software. |
| **open–loop-system** | A system in which controlling action or input is independent of the output or changes in output. |
| **operability** | The degree to which a component or system has attributes that make it easy to operate and control. |
| **operational acceptance testing** production acceptance testing | A type of acceptance testing performed to determine if operations and/or systems administration staff can accept a system. |
| **operational profile** | An actual or predicted pattern of use of the component or system. |
| **operational profiling** | The process of developing and implementing an operational profile. |
| **outsourced testing** | Testing performed by people who are not co-located with the project team and are not fellow employees. |
| **pair testing** | An approach in which two team members simultaneously collaborate on testing a work product. |
| **pairwise integration testing** | A type of integration testing that targets pairs of components that work together as shown in a call graph. |
| **pairwise testing** | A black-box test technique in which test cases are designed to exercise pairs of parameter-value pairs. |
| **par sheet testing** | Testing to determine that the game returns the correct mathematical results to the screen, to the players' accounts, and to the casino account. |
| **Pareto analysis** | A statistical technique in decision making that is used for selection of a limited number of factors that produce significant overall effect. In terms of quality improvement, a large majority of problems (80%) are produced by a few key causes (20%). |
| **pass/fail criteria** | Decision rules used to determine whether a test item has passed or failed. |
| **passed** | The status of a test result in which the actual result matches the expected result. |
| **password cracking** | A security attack recovering secret passwords stored in a computer system or transmitted over a network. |
| **Path** control flow path | A sequence of consecutive edges in a directed graph. |
| **path coverage** | The coverage of paths. |
| **path testing** | A white-box test technique in which test cases are designed to execute paths in a control flow graph. |
| **peer review** | A review performed by others with the same abilities to create the work product. |
| **penetration testing** | A testing technique aiming to exploit security vulnerabilities (known or unknown) to gain unauthorized access. |
| **performance efficiency** | The degree to which a component or system uses time, resources and capacity when accomplishing its designated functions. |
| **performance indicator** key performance indicator | A metric that supports the judgment of process performance. |
| **performance testing** | Testing to determine the performance efficiency of a component or system. |
| **performance testing tool** | A test tool that generates load for a designated test item and that measures and records its performance during test execution. |
| **perspective-based Reading** perspective-based reviewing | A review technique in which a work product is evaluated from the perspective of different stakeholders with the purpose to derive other work products. |
| **pharming** | A security attack intended to redirect a web site's traffic to a fraudulent web site without the user's knowledge or consent. |
| **phase containment** | The percentage of defects that are removed in the same phase of the software lifecycle in which they were introduced. |
| **phishing** | An attempt to acquire personal or sensitive information by masquerading as a trustworthy entity in an electronic communication. |
| **planning poker** | A consensus-based estimation technique, mostly used to estimate effort or relative size of user stories in Agile software development. It is a variation of the Wideband |
| | Delphi method using a deck of cards with values representing the units in which the team estimates. |
| **player perspective testing** | Testing done by testers from a player's perspective to validate player satisfaction. |
| **pointer** | A data item that specifies the location of another data item. |
| **portability** | The degree to which a component or system can be transferred from one hardware, software or other operational or usage environment to another. |
| **post-release testing** | A type of testing to ensure that the release is performed correctly and the application can be deployed. |
| **postcondition** | The expected state of a test item and its environment at the end of test case execution. |
| **precondition** | The required state of a test item and its environment prior to test case execution. |
| **priority** | The level of (business) importance assigned to an item, e.g., defect. |
| **PRISMA** | A systematic approach to risk-based testing that employs product risk identification and analysis to create a product risk matrix based on likelihood and impact. Term is derived from Product RISk MAnagement. |
| **probe effect** | An unintended change in behavior of a component or system caused by measuring it. |
| **process assessment** | A disciplined evaluation of an organization's software processes against a reference model. |
| **process model** | A framework in which processes of the same nature are classified into an overall model. |
| **process-compliant test strategy** | A test strategy whereby the test team follows a set of predefined processes, whereby the processes address such items as documentation, the proper identification and use of the test basis and test oracle(s), and the organization of the test team. |
| **process-driven scripting** | A scripting technique where scripts are structured into scenarios which represent use cases of the software under test. The scripts can be parameterized with test data. |
| **product risk** | A risk impacting the quality of a product. |
| **product-based quality** | A view of quality, wherein quality is based on a well-defined set of quality characteristics. These characteristics must be measured in an objective and quantitative way. Differences in the quality of products of the same type can be traced back to the way the specific quality characteristics have been implemented. |
| **project risk** | A risk that impacts project success. |
| **proximity-based testing** | A type of testing to confirm that sensors can detect nearby objects without physical contact. |
| **quality** | The degree to which a component or system satisfies the stated and implied needs of its various stakeholders. |
| **quality assurance (QA)** | Activities focused on providing confidence that quality requirements will be fulfilled. |
| **quality characteristic** quality attribute; software quality characteristic; software product characteristic | A category of quality attributes that bears on work product quality. |
| **quality control (QC)** | A set of activities designed to evaluate the quality of a component or system. |
| **quality function deployment (QFD)** | A facilitated workshop technique that helps determine critical characteristics for new product development. |
| **quality management** | Coordinated activities to direct and control an organization with regard to quality that include establishing a quality policy and quality objectives, quality planning, quality control, quality assurance, and quality improvement. |
| **quality risk** | A product risk related to a quality characteristic. |
| **RACI matrix** | A matrix describing the participation by various roles in completing tasks or deliverables for a project or process. It is especially useful in clarifying roles and responsibilities. RACI is an acronym derived from the four key responsibilities most typically used: Responsible, Accountable, Consulted, and Informed. |
| **ramp-down** | A technique for decreasing the load on a system in a measurable and controlled way. |
| **ramp-up** | A technique for increasing the load on a system in a measurable and controlled way. |
| **random testing** | A black-box test technique in which test cases are designed by generating random independent inputs to match an operational profile. |
| **Rational Unified Process (RUP)** | A proprietary adaptable iterative software development process framework consisting of four project lifecycle phases: inception, elaboration, construction and transition. |
| **reactive test strategy** | A test strategy whereby the test team waits to design and implement tests until the software is received, reacting to the actual system under test. |
| **reactive testing** | Testing that dynamically responds to the behavior of the test object and to test results being obtained. |
| **Reconnaissance**<br>footprinting | The exploration of a target area aiming to gain information that can be useful for an attack. |
| **recoverability** | The degree to which a component or system can recover the data directly affected by an interruption or a failure and re-establish the desired state of the component or system. |
| **regression testing** | A type of change-related testing to detect whether defects have been introduced or uncovered in unchanged areas of the software. |
| **regression-averse test strategy** | A test strategy whereby the test team applies various techniques to manage the risk of regression such as functional and/or non-functional regression test automation at one or more levels. |
| **regulatory acceptance testing** | A type of acceptance testing performed to verify whether a system conforms to relevant laws, policies and regulations. |
| **reliability** | The degree to which a component or system performs specified functions under specified conditions for a specified period of time. |
| **reliability growth model** | A model that shows the growth in reliability over time of a component or system as a result of the defect removal. |
| **remote test lab** | A facility that provides remote access to a test environment. |
| **replaceability** | The degree to which a component or system can replace another specified component or system for the same purpose in the same environment. |
| **requirement** | A provision that contains criteria to be fulfilled. |
| **requirements-based testing** | An approach to testing in which test cases are designed based on requirements. |
| **resource utilization**<br>storage | The degree to which the amounts and types of resources used by a component or system, when performing its functions, meet requirements. |
| **retrospective meeting**<br>post-project meeting; project retrospective | A meeting at the end of a project during which the project team members evaluate the project and learn lessons that can be applied to the next project. |
| **reusability** | The degree to which a work product can be used in more than one system, or in building other work products. |
| **review** | A type of static testing in which a work product or process is evaluated by one or more individuals to detect defects or to provide improvements. |
| **review plan** | A document describing the approach, resources and schedule of intended review activities. It identifies, amongst others: documents and code to be reviewed, review types to be used, participants, as well as entry and exit criteria to be applied in case of formal reviews, and the rationale for their choice. It is a record of the review planning process. |
| **Reviewer**<br>Checker;<br>inspector | A participant in a review who identifies issues in the work product. |
| **risk** | A factor that could result in future negative consequences. |
| **risk analysis** | The overall process of risk identification and risk assessment. |
| **risk assessment** | The process to examine identified risks and determine the risk level. |
| **risk identification** | The process of finding, recognizing and describing risks. |
| **risk impact**<br>impact | The damage that will be caused if the risk becomes an actual outcome or event. |
| **risk level**<br>risk exposure | The qualitative or quantitative measure of a risk defined by impact and likelihood. |
| **risk likelihood**<br>likelihood | The probability that a risk will become an actual outcome or event. |
| **risk management** | The process for handling risks. |
| **risk mitigation** | The process through which decisions are reached and protective measures are implemented for reducing or maintaining risks to specified levels. |
| **risk type**<br>risk category | A set of risks grouped by one or more common factors. |
| **risk-based testing** | Testing in which the management, selection, prioritization, and use of testing activities and resources are based on corresponding risk types and risk levels. |
| **robustness** | The degree to which a component or system can function correctly in the presence of invalid inputs or stressful environmental conditions. |
| **role-based reviewing** | A review technique in which a work product is evaluated from the perspective of different stakeholders. |
| **root cause** | A source of a defect such that if it is removed, the occurrence of the defect type is decreased or removed. |
| **root cause analysis**<br>causal analysis | An analysis technique aimed at identifying the root causes of defects. By directing corrective measures at root causes, it is hoped that the likelihood of defect recurrence will be minimized. |
| **S.M.A.R.T. goal methodology (SMART)** | A methodology whereby objectives are defined very specifically rather than generically. SMART is an acronym derived from the attributes of the objective to be defined: Specific, Measurable, Attainable, Relevant and Timely. |
| **salting** | A cryptographic technique that adds random data (salt) to the user data prior to hashing. |
| **scalability** | The degree to which a component or system can be adjusted for changing capacity. |
| **scalability testing** | Testing to determine the scalability of the software product. |
| **scenario-based review** | A review technique in which a work product is evaluated to determine its ability to address specific scenarios. |
| **Scribe**<br>recorder | A person who records information during the review meetings. |
| **script kiddie** | A person who executes security attacks that have been created by other hackers rather than creating one's own attacks. |
| **scripted testing** | Testing (manual or automated) that follows a test script. |
| **scrum** | An iterative incremental framework for managing projects commonly used with Agile software development. |
| **security**<br>information security | The degree to which a component or system protects information and data so that persons or other components or systems have the degree of access appropriate to their types and levels of authorization. |
| **security attack** | An attempt to gain unauthorized access to a component or system, resources, information, or an attempt to compromise system integrity. |
| **security audit** | An audit evaluating an organization's security processes and infrastructure. |
| **security policy** | A high-level document describing the principles, approach and major objectives of the organization regarding security. |
| **security procedure** | A set of steps required to implement the security policy and the steps to be taken in response to a security incident. |
| **security risk** | A quality risk related to security. |
| **security testing** | Testing to determine the security of the software product. |
| **security vulnerability** | A weakness in the system that could allow for a successful security attack. |
| **sequential development model** | A type of software development lifecycle model in which a complete system is developed in a linear way of several discrete and successive phases with no overlap between them. |
| **service virtualization** | A technique to enable virtual delivery of services which are deployed, accessed and managed remotely. |
| **session-based test management (SBTM)** | A method for measuring and managing session-based testing. |
| **session-based testing** | An approach in which test activities are planned as test sessions. |
| **severity** | The degree of impact that a defect has on the development or operation of a component or system. |
| **short-circuiting** | A programming language/interpreter technique for evaluating compound conditions in which a condition on one side of a logical operator may not be evaluated if the condition on the other side is sufficient to determine the final outcome. |
| **simulator** | A device, computer program or system used during testing, which behaves or operates like a given system when provided with a set of controlled inputs. |
| **smoke test**<br>sanity test;<br>intake test;<br>confidence test | A test suite that covers the main functionality of a component or system to determine whether it works properly before planned testing begins. |
| **social engineering** | An attempt to trick someone into revealing information (e.g., a password) that can be used to attack systems or networks. |
| **software development lifecycle (SDLC)** | The activities performed at each stage in software development, and how they relate to one another logically and chronologically. |
| **software in the loop (SiL)** | Dynamic testing performed using real software in a simulated environment or with experimental hardware. |
| **software lifecycle** | The period of time that begins when a software product is conceived and ends when the software is no longer available for use. The software lifecycle typically includes a concept phase, requirements phase, design phase, implementation phase, test phase, installation and checkout phase, operation and maintenance phase, and sometimes, retirement phase. Note these phases may overlap or be performed iteratively. |
| **software process improvement (SPI)** | A program of activities designed to improve the performance and maturity of the organization's software processes and the results of such a program. |
| **software qualification test** | Testing performed on completed, integrated software to provide evidence for compliance with software requirements. |
| **Software Usability Measurement Inventory (SUMI)** | A questionnaire-based usability testing tool that measures and benchmarks user experience. |
| **specification by example (SBE)** | A development technique in which the specification is defined by examples. |
| **spike testing** | Testing to determine the ability of a system to recover from sudden bursts of peak loads and return to a steady state. |
| **SQL injection** | A type of code injection in the structured query language (SQL). |
| **staged representation** | A model structure wherein attaining the goals of a set of process areas establishes a maturity level; each level builds a foundation for subsequent levels. |
| **standard** | Formal, possibly mandatory, set of requirements developed and used to prescribe consistent approaches to the way of working or to provide guidelines (e.g., ISO/IEC standards, IEEE standards, and organizational standards). |
| **standard-compliant test strategy** | A test strategy whereby the test team follows a standard. Standards followed may be valid e.g., for a country (legislation standards), a business domain (domain standards), or internally (organizational standards). |
| **state transition testing**<br>finite state testing | A black-box test technique in which test cases are designed to exercise elements of a state transition model. |
| **Statement**<br>source statement | An entity in a programming language, which is typically the smallest indivisible unit of execution. |
| **statement coverage** | The coverage of executable statements. |
| **statement testing** | A white-box test technique in which test cases are designed to execute statements. |
| **static analysis** | The process of evaluating a component or system without executing it, based on its form, structure, content, or documentation. |
| **static analyzer**<br>static analysis tool;<br>analyzer | A tool that carries out static analysis. |
| **static code analysis** | The analysis of source code carried out without execution of that software. |
| **static testing** | Testing a work product without the work product code being executed. |
| **stress testing** | A type of performance testing conducted to evaluate a system or component at or beyond the limits of its anticipated or specified workloads, or with reduced availability of resources such as access to memory or servers. |
| **structural coverage** | Coverage measures based on the internal structure of a component or system. |
| **structured scripting** | A scripting technique that builds and utilizes a library of reusable (parts of) scripts. |
| **stub** | A skeletal or special-purpose implementation of a software component, used to develop or test a component that calls or is otherwise dependent on it. It replaces a called component. |
| **suitability testing** | Testing to determine the suitability of a software product. |
| **summative evaluation** | A type of evaluation designed and used to gather conclusions about the quality of a component or system, especially when a substantial part of it has completed design. |
| **system hardening** | The step-by-step process of reducing the security vulnerabilities of a system by applying a security policy and different layers of protection. |
| **system integration testing** | A test level that focuses on interactions between systems. |
| **system of systems** multi-system | Multiple heterogeneous, distributed systems that are embedded in networks at multiple levels and in multiple interconnected domains, addressing largescale inter-disciplinary common problems and purposes, usually without a common management structure. |
| **system qualification test** | Testing performed on the completed, integrated system of software components, hardware components, and mechanics to provide evidence for compliance with system requirements and that the complete system is ready for delivery. |
| **system testing** | A test level that focuses on verifying that a system as a whole meets specified requirements. |
| **system throughput** | The amount of data passing through a component or system in a given time period. |
| **system under test (SUT)** | A type of test object that is a system. |
| **System Usability Scale (SUS)** | A simple, ten-item attitude scale giving a global view of subjective assessments of usability. |
| **Systematic Test and Evaluation Process (STEP)** | A structured testing methodology also used as a content-based model for improving the testing process. It does not require that improvements occur in a specific order. |
| **technical review** | A type of formal review by a team of technically-qualified personnel that examines the quality of a work product and identifies discrepancies from specifications and standards. |
| **test** | A set of one or more test cases. |
| **test adaptation layer** | The layer in a test automation architecture which provides the necessary code to adapt test scripts on an abstract level to the various components, configuration or interfaces of the SUT. |
| **test analysis** | The activity that identifies test conditions by analyzing the test basis. |
| **test approach** | The implementation of the test strategy for a specific project. |
| **test architect** | (1) A person who provides guidance and strategic direction for a test organization and for its relationship with other disciplines. (2) A person who defines the way testing is structured for a given system, including topics such as test tools and test data management. |
| **test automation** | The use of software to perform or support test activities. |
| **test automation architecture** | An instantiation of the generic test automation architecture to define the architecture of a test automation solution, i.e., its layers, components, services and interfaces. |
| **test automation engineer** | A person who is responsible for the design, implementation and maintenance of a test automation architecture as well as the technical evolution of the resulting test automation solution. |
| **test automation framework** | A tool that provides an environment for test automation. It usually includes a test harness and test libraries. |
| **test automation manager** | A person who is responsible for the planning and supervision of the development and evolution of a test automation solution. |
| **test automation solution** | A realization/implementation of a test automation architecture, i.e., a combination of components implementing a specific test automation assignment. The components may include commercial off-the-shelf test tools, test automation frameworks, as well as test hardware. |
| **test automation strategy** | A high-level plan to achieve long-term objectives of test automation under given boundary conditions. |
| **test basis** | The body of knowledge used as the basis for test analysis and design. |
| **test case** | A set of preconditions, inputs, actions (where applicable), expected results and postconditions, developed based on test conditions. |
| **test case explosion** | The disproportionate growth of the number of test cases with growing size of the test basis, when using a certain test design technique. Test case explosion may also happen when applying the test design technique systematically for the first time. |
| **test case specification** | Documentation of a set of one or more test cases. |
| **test charter**<br>charter | Documentation of the goal or objective for a test session. |
| **test closure** | During the test closure phase of a test process data is collected from completed activities to consolidate experience, testware, facts and numbers. The test closure phase consists of finalizing and archiving the testware and evaluating the test process, including preparation of a test evaluation report. |
| **test completion** | The activity that makes testware available for later use, leaves test environments in a satisfactory condition and communicates the results of testing to relevant stakeholders. |
| **test condition**<br>test situation;<br>test requirement | A testable aspect of a component or system identified as a basis for testing. |
| **test control** | The activity that develops and applies corrective actions to get a test project on track when it deviates from what was planned. |
| **test cycle** | An instance of the test process against a single identifiable version of the test object. |
| **test data** | Data needed for test execution. |
| **test data preparation** | The activity to select data from existing databases or create, generate, manipulate and edit data for testing. |
| **test data preparation tool**<br>test generator | A type of test tool that enables data to be selected from existing databases or created, generated, manipulated and edited for use in testing. |
| **test definition layer** | The layer in a generic test automation architecture which supports test implementation by supporting the definition of test suites and/or test cases, e.g., by offering templates or guidelines. |
| **test design** | The activity that derives and specifies test cases from test conditions. |
| **test design specification** | Documentation specifying the features to be tested and their corresponding test conditions. |
| **test director** | A senior manager who manages test managers. |
| **test environment**<br>test bed;<br>test rig | An environment containing hardware, instrumentation, simulators, software tools, and other support elements needed to conduct a test. |
| **test estimation** | An approximation related to various aspects of testing. |
| **test execution** | The activity that runs a test on a component or system producing actual results. |
| **test execution automation** | The use of software, e.g., capture/playback tools, to control the execution of tests, the comparison of actual results to expected results, the setting up of test preconditions, and other test control and reporting functions. |
| **test execution layer** | The layer in a generic test automation architecture which supports the execution of test suites and/or test cases. |
| **test execution schedule** | A schedule for the execution of test suites within a test cycle. |
| **test execution tool** | A test tool that executes tests against a designated test item and evaluates the outcomes against expected results and postconditions. |
| **test generation layer** | The layer in a generic test automation architecture which supports manual or automated design of test suites and/or test cases. |
| **test harness** | A collection of stubs and drivers needed to execute a test suite |
| **test hook** | A customized software interface that enables automated testing of a test object. |
| **test implementation** | The activity that prepares the testware needed for test execution based on test analysis and design. |
| **test improvement plan** | A plan for achieving organizational test process improvement objectives based on a thorough understanding of the current strengths and weaknesses of the organization's test processes and test process assets. |
| **test infrastructure** | The organizational artifacts needed to perform testing, consisting of test environments, test tools, office environment and procedures. |
| **test item** | A part of a test object used in the test process. |
| **test leader**<br>lead tester | On large projects, the person who reports to the test manager and is responsible for project management of a particular test level or a particular set of testing activities. |
| **test level**<br>test stage | A specific instantiation of a test process. |
| **test log**<br>test record;<br>test run log | A chronological record of relevant details about the execution of tests. |
| **test logging**<br>test recording | The activity of creating a test log. |
| **test management** | The planning, scheduling, estimating, monitoring, reporting, control and completion of test activities. |
| **test management tool** | A tool that supports test management. |
| **test manager** | The person responsible for project management of testing activities, resources, and evaluation of a test object. |
| **Test Maturity Model integration (TMMi)** | A five-level staged framework for test process improvement, related to the Capability Maturity Model Integration (CMMI), that describes the key elements of an effective test process. |
| **test mission** | The purpose of testing for an organization, often documented as part of the test policy. |
| **test model** | A model describing testware that is used for testing a component or a system under test. |
| **test monitoring** | The activity that checks the status of testing activities, identifies any variances from planned or expected, and reports status to stakeholders. |
| **test object** | The work product to be tested. |
| **test objective** | The reason or purpose of testing. |
| **test Oracle**<br>oracle | A source to determine an expected result to compare with the actual result of the system under test. |
| **test phase** | A distinct set of test activities collected into a manageable phase of a project, e.g., the execution activities of a test level. |
| **test plan** | Documentation describing the test objectives to be achieved and the means and the schedule for achieving them, organized to coordinate testing activities. |
| **test planning** | The activity of establishing or updating a test plan. |
| **Test Point Analysis (TPA)** | A formula based test estimation method based on function point analysis. |
| **test policy**<br>organizational test policy | A high-level document describing the principles, approach and major objectives of the organization regarding testing. |
| **test procedure** | A sequence of test cases in execution order, and any associated actions that may be required to set up the initial preconditions and any wrap up activities post execution. |
| **test process** | The set of interrelated activities comprising of test planning, test monitoring and control, test analysis, test design, test implementation, test execution, and test completion. |
| **test process group (TPG)** | A collection of (test) specialists who facilitate the definition, maintenance, and improvement of the test processes used by an organization. |
| **test process improvement** | A program of activities undertaken to improve the performance and maturity of the organization's test processes. |
| **test process improvement manifesto** | A statement that echoes the Agile manifesto, and defines values for improving the testing process. The values are: flexibility over detailed processes, best practices over templates, deployment orientation over process orientation, peer reviews over quality assurance (departments), business driven over model-driven. |
| **test process improver** | A person implementing improvements in the test process based on a test improvement plan. |
| **test progress report**<br>test status report | A type of test report produced at regular intervals about the progress of test activities against a baseline, risks, and alternatives requiring a decision. |
| **test pyramid** | A graphical model representing the relationship of the amount of testing per level, with more at the bottom than at the top. |
| **test report** | Documentation summarizing test activities and results. |
| **test reporting** | Collecting and analyzing data from testing activities and subsequently consolidating the data in a report to inform stakeholders. |
| **test result**<br>outcome;<br>test outcome;<br>result | The consequence/outcome of the execution of a test. |
| **test run** | The execution of a test suite on a specific version of the test object. |
| **test schedule** | A list of activities, tasks or events of the test process, identifying their intended start and finish dates and/or times, and interdependencies. |
| **test script** | A sequence of instructions for the execution of a test. |
| **test selection criteria** | The criteria used to guide the generation of test cases or to select test cases in order to limit the size of a test. |
| **test session** | An uninterrupted period of time spent in executing tests. |
| **test specification** | The complete documentation of the test design, test cases, and test scripts for a specific test item. |
| **test strategy**<br>organizational test strategy | Documentation aligned with the test policy that describes the generic requirements for testing and details how to perform testing within an organization. |
| **test suite**<br>test set;<br>test case suite | A set of test scripts or test procedures to be executed in a specific test run. |
| **test summary report** | A type of test report produced at completion milestones that provides an evaluation of the corresponding test items against exit criteria. |
| **test technique**<br>test design technique;<br>test specification technique;<br>test case design technique | A procedure used to define test conditions, design test cases, and specify test data. |
| **test tool** | Software or hardware that supports one or more test activities. |
| **test type** | A group of test activities based on specific test objectives aimed at specific characteristics of a component or system. |
| **test-driven development (TDD)** | A software development technique in which the test cases are developed, and often automated, and then the software is developed incrementally to pass those test cases. |
| **test-first approach** | An approach to software development in which the test cases are designed and implemented before the associated component or system is developed. |
| **testability** | The degree to which test conditions can be established for a component or system, and tests can be performed to determine whether those test conditions have been met. |
| **tester** | A person who performs testing. |
| **testing** | The process consisting of all lifecycle activities, both static and dynamic, concerned with planning, preparation and evaluation of a component or system and related work products to determine that they satisfy specified requirements, to demonstrate that they are fit for purpose and to detect defects. |
| **testing quadrants** | A classification model of test types/levels in four quadrants, relating them to two dimensions of test goals: supporting the team vs. critiquing the product, and technology-facing vs. business-facing. |
| **testware** | Work products produced during the test process for use in planning, designing, executing, evaluating and reporting on testing. |
| **think aloud usability testing** | A usability testing technique where test participants share their thoughts with the moderator and observers by thinking aloud while they solve usability test tasks. Think aloud is useful to understand the test participant. |
| **think time** | The amount of time required by a user to determine and execute the next action in a sequence of actions. |
| **time behavior** | The degree to which a component or system can perform its required functions within required response times, processing times and throughput rates. |
| **Total Quality Management (TQM)** | An organization-wide management approach centered on quality, based on the participation of all members of the organization and aiming at long-term success through customer satisfaction, and benefits to all members of the organization and to society. Total Quality Management consists of planning, organizing, directing, control, and assurance. |
| **tour** | A set of exploratory tests organized around a special focus. |
| **TPI Next** | A continuous business-driven framework for test process improvement that describes the key elements of an effective and efficient test process. |
| **traceability** | The degree to which a relationship can be established between two or more work products. |
| **traceability matrix** | A two-dimensional table, which correlates two entities (e.g., requirements and test cases). The table allows tracing back and forth the links of one entity to the other, thus enabling the determination of coverage achieved and the assessment of impact of proposed changes. |
| **transactional analysis** | The analysis of transactions between people and within people's minds; a transaction is defined as a stimulus plus a response. Transactions take place between people and between the ego states (personality segments) within one person's mind. |
| **transcendent-based quality** | A view of quality, wherein quality cannot be precisely defined, but we know it when we see it, or are aware of its absence when it is missing. Quality depends on the perception and affective feelings of an individual or group of individuals toward a product. |
| **unit test framework** | A tool that provides an environment for unit or component testing in which a component can be tested in isolation or with suitable stubs and drivers. It also provides other support for the developer, such as debugging capabilities. |
| **usability** | The degree to which a component or system can be used by specified users to achieve specified goals in a specified context of use. |
| **usability evaluation** | A process through which information about the usability of a system is gathered in order to improve the system (known as formative evaluation) or to assess the merit or worth of a system (known as summative evaluation). |
| **usability lab** | A test facility in which unintrusive observation of participant reactions and responses to software takes place. |
| **usability requirement** | A requirement on the usability of a component or system. |
| **usability test participant** | A representative user who solves typical tasks in a usability test. |
| **usability test script** | A document specifying a sequence of actions for the execution of a usability test. It is used by the moderator to keep track of briefing and pre-session interview questions, usability test tasks, and post-session interview questions. |
| **usability test session** | A test session in usability testing in which a usability test participant is executing tests, moderated by a moderator and observed by a number of observers. |
| **usability test task** | A usability test execution activity specified by the moderator that needs to be accomplished by a usability test participant within a given period of time. |
| **usability testing** | Testing to evaluate the degree to which the system can be used by specified users with effectiveness, efficiency and satisfaction in a specified context of use. |
| **use case testing**<br>user scenario testing; scenario testing | A black-box test technique in which test cases are designed to exercise use case behaviors. |
| **user acceptance testing (UAT)** | A type of acceptance testing performed to determine if intended users accept the system. |
| **user error protection** | The degree to which a component or system protects users against making errors. |
| **user experience** | A person's perceptions and responses resulting from the use or anticipated use of a software product. |
| **user interface** | All components of a system that provide information and controls for the user to accomplish specific tasks with the system. |
| **user interface aesthetics**<br>attractiveness | The degree to which a user interface enables pleasing and satisfying interaction for the user. |
| **user interface guideline** | A low-level, specific rule or recommendation for user interface design that leaves little room for interpretation so designers implement it similarly. It is often used to ensure consistency in the appearance and behavior of the user interface of the systems produced by an organization. |
| **user story** | A user or business requirement consisting of one sentence expressed in the everyday or business language which is capturing the functionality a user needs, the reason behind it, any non-functional criteria, and also including acceptance criteria. |
| **user story testing** | A black-box test design technique in which test cases are designed based on user stories to verify their correct implementation. |
| **user survey** | A usability evaluation whereby a representative sample of users are asked to report subjective evaluation into a questionnaire based on their experience in using a component or system. |
| **user-agent based testing** | A type of testing in which a test client is used to switch the user agent string and identify itself as a different client while executing test suites. |
| **user-based quality** | A view of quality, wherein quality is the capacity to satisfy needs, wants and desires of the user(s). A product or service that does not fulfill user needs is unlikely to find any users. This is a context dependent, contingent approach to quality since different business characteristics require different qualities of a product. |
| **V-model** | A sequential development lifecycle model describing a one-for-one relationship between major phases of software development from business requirements specification to delivery, and corresponding test levels from acceptance testing to component testing. |
| **validation** | Confirmation by examination and through provision of objective evidence that the requirements for a specific intended use or application have been fulfilled. |
| **value-based quality** | A view of quality wherein quality is defined by price. A quality product or service is one that provides desired performance at an acceptable cost. Quality is determined by means of a decision process with stakeholders on trade-offs between time, effort and cost aspects. |
| **verification** | Confirmation by examination and through provision of objective evidence that specified requirements have been fulfilled. |
| **virtual user** | A simulation of activities performed according to a user operational profile. |
| **vulnerability scanner** | A static analyzer that is used to detect particular security vulnerabilities in the code. |
| **Walkthrough**<br>structured walkthrough | A type of review in which an author leads members of the review through a work product and the members ask questions and make comments about possible issues. |
| **Web Content Accessibility Guidelines (WCAG)** | A part of a series of web accessibility guidelines published by the Web Accessibility Initiative (WAI) of the World Wide Web Consortium (W3C), the main international standards organization for the internet. They consist of a set of guidelines for making content accessible, primarily for people with disabilities. |
| **Website Analysis and Measurement Inventory (WAMMI)** | A commercial website analysis service providing a questionnaire for measuring user experience and assessing delivery of business goals online. |
| **white-box test technique**<br>structure-based technique;<br>structure-based test technique;<br>structural test technique;<br>white-box technique | A test technique only based on the internal structure of a component or system. |
| **white-box testing**<br>clear-box testing;<br>code-based testing;<br>glass-box testing;<br>logic-coverage testing;<br>logic-driven testing; | Testing based on an analysis of the internal structure of the component or system. |
| structural testing; structure-based testing | |
| **Wideband Delphi** | An expert-based test estimation technique that aims at making an accurate estimation using the collective wisdom of the team members. |
| **wild pointer** | A pointer that references a location that is out of scope for that pointer or that does not exist. |
| **XiL test environment (XiL)** | A generalized term for dynamic testing in different virtual test environments. |

# Appendix V: Quantities related to the number of terms for each category in the glossaries

*In this Appendix we show the metrics and their obtained values related to the number of terms for each category in the 3 glossaries. We include metrics for the total number of terms for categories 1 to 6, which are specific for the software testing domain, in addition to the total number of terms ending with the word "testing" per category as well.*

| Metric name/acronym | Value |
|---|---|
| Total Sum of Unique Terms in Category1 [TUTC1 = (#UTxISOC1 + #UTxTMMiC1 + #UTxISTQBC1)] | 64 |
| Total Sum of Unique Terms in Category2 [TUTC2 = (#UTxISOC2 + #UTxTMMiC2 + #UTxISTQBC2)] | 191 |
| Total Sum of Unique Terms in Category3 [TUTC3 = (#UTxISOC3 + #UTxTMMiC3 + #UTxISTQBC3)] | 24 |
| Total Sum of Unique Terms in Category4 [TUTC4 = (#UTxISOC4 + #UTxTMMiC4 + #UTxISTQBC4)] | 130 |
| Total Sum of Unique Terms in Category5 [TUTC5 = (#UTxISOC5 + #UTxTMMiC5 + #UTxISTQBC5)] | 75 |
| Total Sum of Unique Terms in Category6 [TUTC6 = (#UTxISOC6 + #UTxTMMiC6 + #UTxISTQBC6)] | 52 |
| Total Sum of Unique Terms ending with the word "Testing" in Category1 (TUTeTC1) | 13 |
| Total Sum of Unique Terms ending with the word "Testing" in Category2 (TUTeTC2) | 109 |
| Total Sum of Unique Terms ending with the word "Testing" in Category3 (TUTeTC3) | 0 |
| Total Sum of Unique Terms ending with the word "Testing" in Category4 (TUTeTC4) | 0 |
| Total Sum of Unique Terms ending with the word "Testing" in Category5 (TUTeTC5) | 32 |
| Total Sum of Unique Terms ending with the word "Testing" in Category6 (TUTeTC6) | 0 |

# Appendix VI: Frequencies and semantic classification of glossaries terms ending with the word "testing" in categories 1, 2 and 5

*In this Appendix we document a set of tables that contain: i) terms with the syntactic frequency of 3; ii) terms with the syntactic frequency of 2; and iii) terms with the syntactic frequency of 1. Note that all term names end with the word "testing". Also, we include the semantic classification of each term into C1 (Cat1_Sgy – Strategy), C2 (Cat2_Wpr – Work Processes) or C5 (Cat5_Me – Methods). Note also that these tables support further analysis of terms with respect to syntactic/semantic similarities/discrepancies and ultimately syntactic/semantic consistencies as carried out in Section 3 of the conference paper* [4].

| **Full Syntactic Similarity (Frequency 3) for terms ending with "testing" in Categories 1 to 6**<br>Note that Cat1_Sgy stands for Category 1 (Strategy), Cat2_WPr for Category 2 (Work Process), and Cat5_Me for Category 5 (Method) | | | | |
|---|---|---|---|---|
| **Term**<br>*Green shaded indicates full semantic similarity<br>*Yellow shaded indicates partial semantic similarity<br>*Red shaded indicates no semantic similarity | **ISO 29119-1** | **TMMi** | **ISTQB** | **Glossaries' main entries and synonyms**<br><br>Note that if the cell is empty, it means that none of the main entries have synonyms |
| testing | Cat2_WPr | Cat2_WPr | Cat2_WPr | |
| dynamic testing | Cat2_WPr | Cat2_WPr | Cat2_WPr | |
| static testing | Cat2_WPr | Cat2_WPr | Cat2_WPr | |
| white-box testing | Cat2_WPr | Cat2_WPr | Cat2_WPr | *ISO 29119-1:<br>Main entry: structure-based testing<br>Synonyms: structural testing; glass-box testing; white box testing<br><br>*TMMi:<br>Main entry: white-box testing<br><br>*ISTQB:<br>Main entry: white-box testing<br>Synonyms: clear-box testing; code-based testing; glass-box testing; logic-coverage testing; logic-driven testing; structural testing; structure-based testing |
| regression testing | Cat2_WPr | Cat2_WPr | Cat2_WPr | |
| confirmation testing | Cat2_WPr | Cat2_WPr | Cat2_WPr | *ISO 29119-1:<br>Main entry: retesting<br>Synonyms: confirmation testing<br><br>*TMMi:<br>Main entry: re-testing<br>Synonyms: confirmation testing |
| | | | | *ISTQB:<br>Main entry: confirmation testing<br>Synonyms: re-testing |
| statement testing | Cat2_WPr | Cat5_Me | Cat5_Me | |
| scenario testing | Cat2_WPr | Cat5_Me | Cat5_Me | *ISO 29119-1:<br>Main entry: scenario testing<br><br>*TMMi:<br>Main entry: use case testing<br><br>*ISTQB:<br>Main entry: use case testing<br>Synonyms: user scenario testing; scenario testing |
| risk-based testing | Cat2_WPr | Cat1_Sgy | Cat2_WPr | |
| exploratory testing | Cat2_WPr | Cat5_Me | Cat1_Sgy | |

| **Partial Syntactic Similarity (Frequency 2) for terms ending with "testing" in Categories 1 to 6**<br>Note that Cat1_Sgy stands for Category 1 (Strategy), Cat2_WPr for Category 2 (Work Process), and Cat5_Me for Category 5 (Method) | | | | |
|---|---|---|---|---|
| **Term**<br>*Green shaded indicates full semantic similarity<br>*Yellow shaded indicates no semantic similarity | **ISO 29119-1** | **TMMi** | **ISTQB** | **Glossaries' main entries and synonyms**<br>Note that if the cell is empty, it means that none of the main entries have synonyms |
| acceptance testing | | Cat2_WPr | Cat2_WPr | |
| accessibility testing | Cat2_WPr | | Cat2_WPr | |
| alpha testing | | Cat2_WPr | Cat2_WPr | |
| state transition testing | | Cat5_Me | Cat5_Me | *TMMi:<br>Main entry: state transition testing<br><br>*ISTQB:<br>Main entry: state transition testing<br>Synonyms: finite state testing |
| exhaustive testing | | Cat1_Sgy | Cat1_Sgy | *TMMi:<br>Main entry: exhaustive testing<br><br>*ISTQB:<br>Main entry: exhaustive testing<br>Synonyms: complete testing |
| independence of testing | | Cat1_Sgy | Cat1_Sgy | |
| capacity testing | Cat2_WPr | | Cat2_WPr | |
| requirements-based testing | | Cat1_Sgy | Cat1_Sgy | |
| security testing | Cat2_WPr | | Cat2_WPr | |
| component integration testing | | Cat2_WPr | Cat2_WPr | *TMMi:<br>Main entry: component integration testing<br><br>*ISTQB:<br>Main entry: component integration testing |
| | | | | Synonyms: link testing |
| component testing | | Cat2_WPr | Cat2_WPr | *TMMi:<br>Main entry: component testing<br>Synonyms: unit testing<br><br>*ISTQB:<br>Main entry: component testing<br>Synonyms: unit testing; module testing |
| beta testing | | Cat2_WPr | Cat2_WPr | |
| stress testing | Cat2_WPr | | Cat2_WPr | |
| decision testing | | Cat5_Me | Cat5_Me | |
| system integration testing | | Cat2_WPr | Cat2_WPr | |
| endurance testing | Cat2_WPr | | Cat2_WPr | |
| system testing | | Cat2_WPr | Cat2_WPr | |
| functional testing | | Cat2_WPr | Cat2_WPr | |
| load testing | Cat2_WPr | | Cat2_WPr | |
| scripted testing | Cat2_WPr | | Cat2_WPr | |
| integration testing | | Cat2_WPr | Cat2_WPr | |
| performance testing | Cat2_WPr | | Cat2_WPr | |
| black-box testing | Cat2_WPr | Cat2_WPr | | *ISO 29119-1:<br>Main entry: specification-based testing<br>Synonyms: black-box testing; closed box testing<br><br>*TMMi:<br>Main entry: black-box testing |
| decision table testing | | Cat5_Me | Cat5_Me | |
| non-functional testing | | Cat2_WPr | Cat2_WPr | |

## Without Syntactic Similarity (Frequency 1) for terms ending with “testing” in Categories 1 to 6

Note that Cat1_Sgy stands for Category 1 (Strategy), Cat2_WPr for Category 2 (Work Process), and Cat5_Me for Category 5 (Method)

| **Term** | **ISO 29119-1** | **TMMi** | **ISTQB** | **Synonyms**<br><br>Note that if the cell is empty it means there are no synonyms |
|---|---|---|---|---|
| agile testing | | | Cat1_Sgy | |
| api testing | | | Cat2_WPr | |
| audio testing | | | Cat2_WPr | |
| computer aided software testing (cast) | | | Cat1_Sgy | |
| back-to-back-testing | | | Cat2_WPr | |
| backup and recovery testing | Cat2_WPr | | | |
| continuous testing | | | Cat1_Sgy | |
| crowd testing | | | Cat1_Sgy | |
| branch testing | | Cat5_Me | | |
| change-related testing | | | Cat2_WPr | |
| checklist-based testing | | | Cat5_Me | |
| cli testing | | | Cat2_WPr | |
| combinatorial testing | | | Cat5_Me | |
| compatibility testing | Cat2_WPr | | | |
| compliance testing | | | Cat2_WPr | Synonyms: conformance testing; regulation testing; standards testing |
| concurrency testing | | | Cat2_WPr | |
| condition testing | | Cat5_Me | | |
| contractual acceptance testing | | | Cat2_WPr | |
| control flow testing | | | Cat5_Me | |
| data-driven testing | | | Cat5_Me | |
| decision condition testing | | | Cat5_Me | |
| device-based testing | | | Cat2_WPr | |
| elementary comparison testing | | Cat5_Me | | |
| end-to-end testing | | | Cat2_WPr | Synonyms: E2E testing |
| experience-based testing | | | Cat2_WPr | |
| field testing | | | Cat2_WPr | |
| fuzz testing | | | Cat5_Me | Synonyms: fuzzing |
| gui testing | | | Cat2_WPr | |
| insourced testing | | | Cat2_WPr | |
| installability testing | Cat2_WPr | | | |
| interface testing | | | Cat2_WPr | |
| interoperability testing | | | Cat2_WPr | |
| keyword-driven testing | | | Cat5_Me | Synonyms: action word-driven testing |
| maintainability testing | Cat2_WPr | | | |
| maintenance testing | | | Cat2_WPr | |
| math testing | | | Cat2_WPr | |
| model-based testing (MBT) | | | Cat2_WPr | |
| modified condition/decision testing | | | Cat5_Me | Synonyms: modified multiple condition testing; condition determination testing |
| multiplayer testing | | | Cat2_WPr | |
| multiple condition testing | | | Cat5_Me | Synonyms: branch condition combination testing; condition combination testing |
| negative testing | | | Cat2_WPr | Synonyms: invalid testing; dirty testing |
| neighborhood integration testing | | | Cat2_WPr | |
| operational acceptance testing | | | Cat2_WPr | Synonyms: production acceptance testing |
| operational profile testing | | Cat2_WPr | | |
| outsourced testing | | | Cat2_WPr | |
| pair testing | | | Cat2_WPr | |
| pairwise integration testing | | | Cat2_WPr | |
| pairwise testing | | | Cat5_Me | |
| par sheet testing | | | Cat2_WPr | |
| path testing | | | Cat5_Me | |
| penetration testing | | | Cat5_Me | |
| player perspective testing | | | Cat2_WPr | |
| portability testing | Cat2_WPr | | | |
| post-release testing | | | Cat2_WPr | |
| procedure testing | Cat2_WPr | | | |
| proximity-based testing | | | Cat2_WPr | |
| random testing | | | Cat5_Me | |
| reactive testing | | | Cat2_WPr | |
| regulatory acceptance testing | | | Cat2_WPr | |
| reliability testing | Cat2_WPr | | | |
| session-based testing | | | Cat1_Sgy | |
| scalability testing | | | Cat2_WPr | |
| spike testing | | | Cat2_WPr | |
| statistical testing | | Cat5_Me | | |
| suitability testing | | | Cat2_WPr | |
| syntax testing | | Cat5_Me | | |
| think aloud usability testing | | | Cat5_Me | |
| unscripted testing | Cat2_WPr | | | |
| usability testing | | | Cat2_WPr | |
| user acceptance testing (UAT) | | | Cat2_WPr | |
| user story testing | | | Cat5_Me | |
| user-agent based testing | | | Cat2_WPr | |
| volume testing | Cat2_WPr | | | |
| equivalence partitioning | | | Cat5_Me | Synonyms: partition testing |